\documentclass[a4paper,11pt]{article}
\usepackage{jinstpub}
\usepackage{lineno}
\usepackage{siunitx}
\title{\boldmath Characterisation of silicon photomultipliers in a dilution refrigerator down to 9.4 mK towards a cryogenic cosmic-ray muon veto system}

\author[]{QUEST-DMC Collaboration:}
\author[a]{A. Kemp,}
\author[b]{S. Autti,}
\author[c]{E. Bloomfield,}
\author[d]{A. Casey,}
\author[d]{N. Darvishi,}
\author[d]{D. Doling,}
\author[d]{N. Eng,}
\author[c]{P. Franchini,}
\author[b]{R. P. Haley,}
\author[d]{P. J. Heikkinen,}
\author[e]{A. Jennings,}
\author[c,d]{S. Koulosousas}
\author[c,d]{E. Leason,}
\author[d]{L. V. Levitin,}
\author[c]{J. March-Russell,}
\author[b]{A. Mayer,}
\author[a,c]{J. Monroe,}
\author[b]{D. M\"unstermann,}
\author[b]{M. T. Noble,}
\author[b]{J. R. Prance,}
\author[d]{X. Rojas,}
\author[b]{T. Salmon,}
\author[d]{J. Saunders,}
\author[f]{J. Smirnov,}
\author[a]{R. Smith,}
\author[b]{M. D. Thompson,}
\author[b]{A. Thomson,}
\author[d]{A. Ting,}
\author[b]{V. Tsepelin,}
\author[d]{S. M. West,}
\author[b]{L. Whitehead,}
\author[b]{D. E. Zmeev.}
\affiliation[a]{UKRI STFC Rutherford Appleton Laboratory, Particle Physics Department, Harwell, Didcot,\\OX11 0QX, UK.}
\affiliation[b]{Department of Physics, Bailrigg, Lancaster University, Lancaster, LA1 4YB, UK.}
\affiliation[c]{Department of Physics, University of Oxford, Keble Road, Oxford, OX1 3RH, UK.}
\affiliation[d]{Department of Physics, Royal Holloway University of London, Egham Hill, Egham, Surrey, TW20 0EX, UK.}
\affiliation[e]{RIKEN Center for Quantum Computing, RIKEN, Wako, 351-0198, Japan.}
\affiliation[f]{Department of Mathematical Sciences, Peach Street, University of Liverpool, Liverpool, L69 7ZL, UK.}

\emailAdd{ashlea.kemp@stfc.ac.uk}

\abstract{We report the characterisation of a FBK NUV-HD-cryo silicon photomultiplier (SiPM) sensor operated in a 9.4 $\pm$ 0.2 mK environment inside a dilution refrigerator, towards the development of a cryogenic cosmic-ray muon veto system to be operated internal to a dilution refrigerator required for low background experiments such as the QUEST-DMC dark matter search experiment. We characterise the single photon response and the gain (the charge produced per detected photon), the dark count noise rate, and correlated noise contributions as a function of operating voltage. This paper also reports first proof-of-concept measurements of using a SiPM coupled to scintillator internal to a dilution refrigerator, towards detecting high-energy events consistent with candidate cosmic-ray muon signals.}


\begin{document}
\maketitle
\flushbottom

\section{Introduction}
\label{sec:Introduction}
In recent years, silicon photomultipliers (SiPMs) have emerged as promising single photon detectors for use in particle physics, particularly in low background, rare-event search experiments such as direct dark matter detection experiments. Photomultiplier Tubes (PMTs) have long been the favoured technology for dark matter experiments, such as DEAP-3600 and LZ~\cite{DEAPPMT,LZPMT}, however compared to PMTs, SiPMs are cheaper to manufacture, require a lower operational voltage, and are insensitive to magnetic fields~\cite{espana2010performance}. In addition to their high gain and high photon detection efficiency, they can have a lower intrinsic radioactivity compared to PMTs. One drawback of SiPMs is that they generally exhibit higher intrinsic thermal noise than PMTs. However, since thermal noise decreases proportionally with temperature, it can be suppressed by operating the SiPMs at cryogenic temperatures. 

Whilst limited, there have been measurements showing that SiPMs can operate close to liquid helium (LHe) temperatures and below~\cite{SiPM4K,hanski2025performance,zhang2022scintillation}, illustrating that there is a growing interest to demonstrate the viability of operating these sensors in sub-Kelvin conditions. As such, SiPMs are being explored by the QUEST-DMC collaboration as a potential photon sensor for use in a cryogenic cosmic-ray muon veto system. The QUEST-DMC collaboration is developing, in one of its work packages, a direct dark matter search experiment projected to reach world-leading sensitivity to spin-dependent sub-GeV/c$^{2}$ mass dark matter using a superfluid helium-3 ($^{3}$He) target enclosed in a $\sim$1 cm$^{3}$ bolometer box~\cite{QUESTsensitivity,QUEST-DMC:2025qsa,QUEST-DMC:2025miz}. The QUEST-DMC collaboration is operating bolometers at Royal Holloway, University of London and Lancaster University, in ultra-low temperature dilution refrigerators fitted with an adiabatic demagnetisation stage capable of cooling $^{3}$He to operating temperatures below 300~\textmu K. 

Since QUEST-DMC is currently operated at sea level, and not underground like the majority of dark matter experiments, cosmic-ray induced background interactions will be a limiting factor for QUEST-DMC's dark matter search~\cite{autti2024quest}. The cosmic-ray muon veto system is currently in development; the baseline design makes use of a scintillator volume surrounding the bolometer cell, where the exterior of this volume is coupled to the inner vacuum can which itself is coupled to the 10 mK stage of the dilution refrigerator, and is where the SiPM will be mounted. Therefore, 10 mK is the relevant temperature to characterise the SiPM in order to demonstrate the feasibility of this scheme. We are considering SiPMs over technologies traditionally deployed for photon detection in this temperature range due to the fact that they are large-area, commercially manufacturable, and simple to operate, compared with Superconducting Nanowire Single-Photon detectors (SNSPDs) which are much smaller, and Transition Edge Sensors (TESs) which have a relatively complex fabrication process.

SiPM technology has previously been used in cryogenic cosmic-ray muon veto systems, such as the NUCLEUS experiment~\cite{NUCLEUS}, which makes use of a plastic scintillator deployed in a dilution refrigerator at 850 mK. In this configuration, the plastic scintillator is instrumented with wavelength shifting fibres, and is coupled to a SiPM that is thermally coupled to the 300 K part of the cryostat. However, we are exploring the option of directly coupling SiPMs to the plastic scintillator itself, which would require the SiPM technology to be able to operate in mK environments. Without amplifier electronics and in the absence of light, a single NUV-HD-cryo SiPM with a surface area of 12 $\times$ 8 mm$^{2}$ dissipates $\mathcal{O}$(pW) of power whilst it is biased, based on the device's properties measured at 77 K~\cite{NUV-HD-cryo}. At this level of power dissipation, the heat load should be low enough such that a SiPM could be thermally coupled to a $\mathcal{O}$(10's) mK temperature stage of a dilution refrigerator without disrupting the overall cooling power of the cryostat.

In this paper, we show the first operation of a NUV-HD-cryo SiPM~\cite{NUV-HD-cryo} in a dilution refrigerator, demonstrating the feasibility of deploying this technology in a cryogenic cosmic-ray muon system for the QUEST-DMC experiment. Section~\ref{sec:exp_setup} outlines the experimental setup, the data acquisition, and data processing chain. Section~\ref{sec:Characterisation} describes the characterisation analysis and results. Section~\ref{sec:cosmicveto} reports first proof-of-principle measurements of using a SiPM coupled to a scintillator and operated internal to a dilution refrigerator. To conclude, Section~\ref{sec:summary} discusses whether, based on the characterisation measurements presented in Section~\ref{sec:Characterisation}, the SiPM tested could meet the requirements of a cryogenic cosmic-ray muon veto system, and how this work informs the next-step R\&D for the system.

\section{Experimental setup}
\label{sec:exp_setup}
\subsection{Hardware}
\begin{figure}[htbp]
\centering
\includegraphics[width=.9\textwidth]{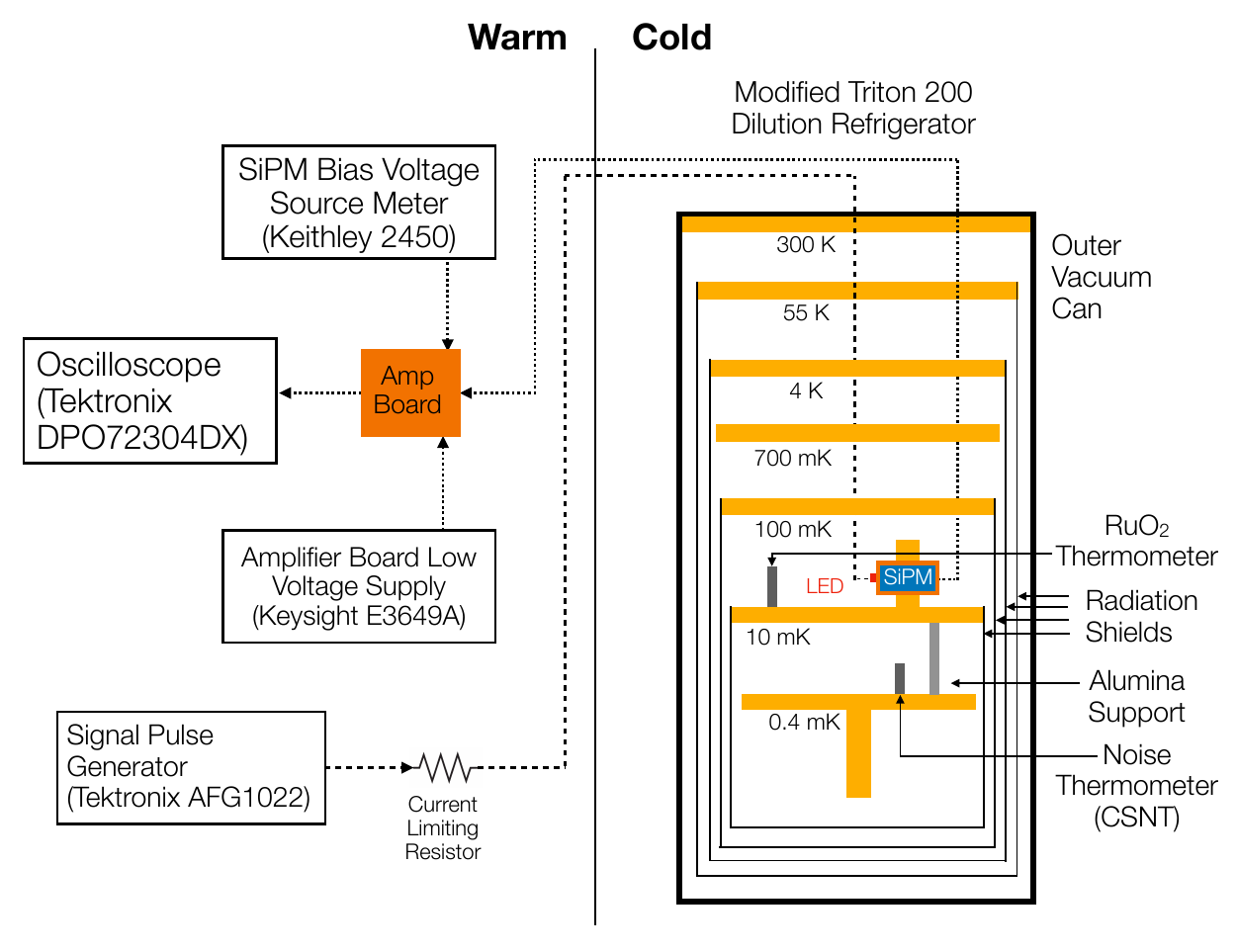}
\caption{Simplified schematic diagram of the experimental setup used for the operation of a single NUV-HD-cryo SiPM in a cryogen-free dilution refrigerator.}
\qquad
\label{F:ExpSetup}
\end{figure}

\begin{figure}[htbp]
\centering
\hspace*{-1cm}  
\includegraphics[width=1.1\textwidth]{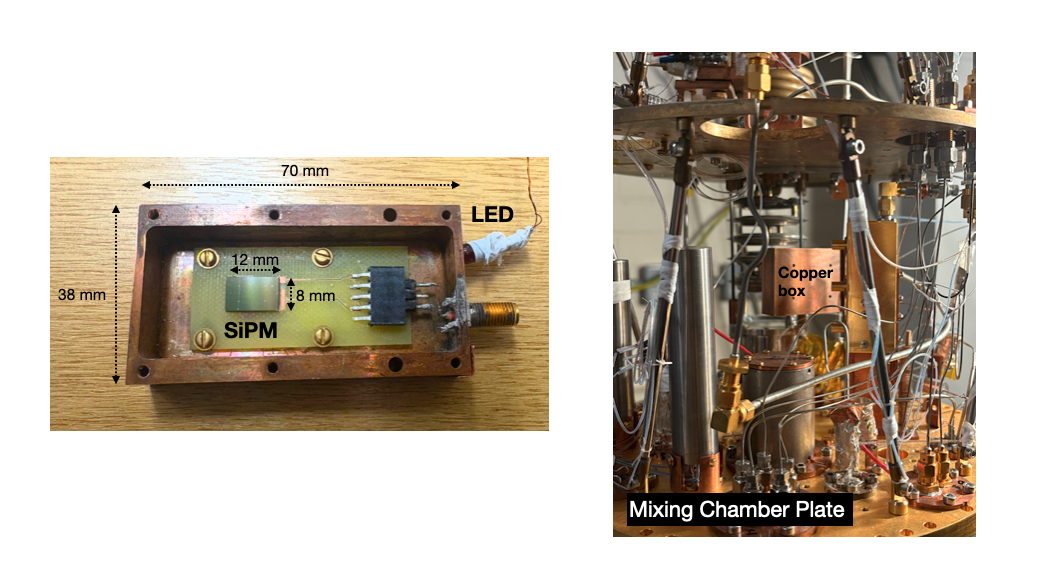}
\caption{Left: Picture of the single NUV-HD-cryo SiPM tested in this work, mounted onto a PCB and housed inside a 7 cm (L) $\times$ 3.8 cm (W) $\times$ 1.4 cm copper box. An LED that emits in the red wavelength range is glued to the outside of the copper box using araldite, over a 1 mm clearance hole allowing direct line of sight to the SiPM for acquiring reverse I-V curves in dark conditions. Right: Picture of the copper box (shown in left figure) mounted inside the dilution refrigerator. The copper box is attached to a copper stand thermally coupled to the mixing chamber plate, which reaches a base temperature of $T_{\mathrm{CSNT}}$ = 9.4 $\pm$ 0.2 mK.}
\qquad
\label{F:SiPMinND4}
\end{figure}

We operate a single NUV-HD-cryo SiPM~\cite{NUV-HD-cryo} of 12 $\times$ 8 mm$^{2}$ surface area and 30 \textmu m cell pitch inside a cryogen-free dilution refrigerator located at Royal Holloway, University of London. The dilution refrigerator is a modified Oxford Instruments Triton 200~\cite{ND4}. A simplified schematic of the experimental setup is shown in Figure~\ref{F:ExpSetup}. The SiPM is mounted onto a custom PCB that distributes bias voltage and is housed inside a 7 cm (L) $\times$ 3.8 cm (W) $\times$ 1.4 cm (D) copper box, shown in Figure~\ref{F:SiPMinND4}. The copper box is mounted and thermally coupled to the mixing chamber plate via a copper stand attached perpendicular to the mixing chamber plate, also shown in Figure~\ref{F:SiPMinND4}. 


The base temperature is measured using current sensing noise thermometry (CSNT)~\cite{CSNT}. A CSNT sensor with resistance 2 m$\Omega$ is mounted onto the ultra-low temperature plate, which is suspended from the mixing chamber plate using alumina tubes. The mixing chamber plate is thermally coupled to the ultra-low temperature plate via a flexible copper link. Whilst the data was acquired for this work, the base temperature as measured from the CSNT sensor on the ultra-low temperature plate, and thus, the mixing chamber plate, was $T_{\mathrm{CSNT}}$ = 9.4 $\pm$ 0.2 mK. Further detail on the fridge design can be found in~\cite{ND4}. To ensure that the characterisation results we obtain actually reflect the SiPM operating in a 9.4 mK environment, we waited at least three days after cool-down to the mixing chamber base temperature before acquiring any data for this work. This allows for the plastic (FR-4) PCB on which the SiPM is mounted to thermalise to the same temperature as the copper. 

We estimate the maximum temperature difference between the SiPM and the mixing chamber temperature measured by the CSNT thermometer assuming the worst possible thermal link, i.e., the SiPM is directly mounted onto the FR-4 PCB which itself is sat directly on the mixing chamber plate  (SiPM $\rightarrow$ FR-4 $\rightarrow$ mixing chamber). In this scenario, all of the heat would flow vertically through the FR-4 bulk only. First, we estimate the power dissipation of the SiPM whilst operated in the dark, $P_{\mathrm{SiPM}}$, as,

\begin{equation}
    P_{\mathrm{SiPM}} = \mathrm{DCR}\cdot G\cdot q\cdot V_{\mathrm{bias}},
\end{equation}
where DCR is the dark count rate, $G$ is the gain, $q = 1.6\times10^{-19}$ C is the electron charge, and $V_{\mathrm{bias}}$ is the bias voltage. We take the upper DCR value reported at 77 K from \cite{NUV-HD-cryo} of 5 mHz/mm$^{2}$. We assume a gain of $3\times10^{6}$, taken also from \cite{NUV-HD-cryo} for a SiPM with 30 \textmu m pitch and at 5 V of overvoltage. We take $V_{\mathrm{bias}}$ = 34 V to correspond with the highest bias voltage we acquired characterisation data. With these values, we calculate $P_{\mathrm{SiPM}}$ = 8.2 pW, which we round up to 10 pW. We also add an additional 10 pW of power attributed to wiring within the dilution fridge, yielding a total $P_{\mathrm{SiPM}}$ = 20 pW. We calculate the temperature difference between the SiPM and mixing chamber, $\Delta T = P_{\mathrm{SiPM}}\cdot R_{\mathrm{th}}$, where $R_{\mathrm{th}}$ is the thermal resistance between the SiPM and mixing chamber. $R_{\mathrm{th}}$ is defined as:

    \begin{equation}
        R_{\mathrm{th}} = \frac{L}{k\cdot A},
    \end{equation}
where $L$ is the PCB thickness = 1.6 mm, $A$ is the SiPM surface area = 96 mm$^{2}$, and $k$ is the thermal conductivity of FR-4. We take $k$ = 1.64 mW/(m$\cdot$K) as measured at 0.3 K in~\cite{runyan2008thermal}. With this, we calculate a worst-case $R_{\mathrm{th}}=1.016\times10^{4}$ K/W. In reality, we anticipate $R_{\mathrm{th}}$ to be much lower owing to additional thermal pathways of: the soldering of the SiPM onto copper pads on the PCB; the metal screws attaching the PCB to the copper box; the electrical grounding to the copper box; and the direct mounting of the copper box onto the flexible copper stand attached to mixing chamber plate. With these values, we make an estimate of $\Delta T$ = 203 nK. This temperature difference is negligible compared to $T_{\mathrm{CSNT}}$ = 9.4 $\pm$ 0.2 mK, thus we are confident that the temperature of SiPM is essentially at the temperature of the mixing chamber. Furthermore, since we are operating at the base temperature of this fridge, there is no more cooling power available at the mixing chamber; even 1 \textmu W of injected power would cause a measurable increase in the CSNT temperature ($\Delta T\sim$ 10 mK assuming the worst-case thermal coupling), which we did not observe.

As Figure~\ref{F:ExpSetup} shows, all associated electronics are located outside of the dilution refrigerator at room temperature, including a custom-designed signal amplification board, providing a 130 V/V voltage gain. A Keysight E3649A provides the low-voltage supply required to power the board. A Keithley 2450 sourcemeter is used to deliver the bias voltage to the SiPM via the custom amplifier board and an external coaxial cable that connects to the top of the cryostat; an internal coaxial cable then delivers the bias voltage down to the SiPM mounted on the mixing chamber plate on the cold side. In addition, an LED that emits in the red wavelength range is attached to the outside of the copper box, where there is a 1 mm clearance hole allowing a direct line of sight from the LED to the SiPM. The LED is connected to a signal pulse generator via a current-limiting resistor, both located outside of the dilution refrigerator, which is used to flash the LED whilst acquiring reverse I-V curves, discussed in Section~\ref{sub:IV}. Data is acquired with a Tektronix DPO72304DX 23 GHz bandwidth oscilloscope~\cite{tetronixscope}.

\subsection{Data acquisition and processing}
\label{subsec:DAQ}
We acquired three long datasets corresponding to three different  operating bias voltages: $V_{\mathrm{bias}} $ = 32 V, 33 V, and 34 V. Each dataset was acquired in dark conditions, i.e., the LED was switched off. In principle, there should be no other external light reaching the SiPM, as this fridge is inherently dark due to the significant radiation shielding implemented to enable reaching ultra-low temperatures. Therefore, the only signals we expect to observe in these datasets should originate from thermally-induced intrinsic noise associated with the SiPM known as the dark count, plus any additional correlated noise sources. The dark count rate of the NUV-HD-cryo technology is expected to be extremely low, as it is measured to be $<$5 mHz/mm$^{2}$ at 77 K~\cite{NUV-HD-cryo}. 

Approximately 300,000 waveforms were acquired in self-trigger mode per voltage. Data were acquired using the Tektronix FastFrame Memory scheme~\cite{FastFrame}, which segments the available memory in the oscilloscope such that each segment can be filled with a triggered acquisition at the desired sample rate with minimal dead time between the acquisitions. This is ideal for applications where pulses are separated by large intervals of inactivity in the signal~\cite{FastFrame}, such as in our case. We acquire waveforms with a 250 MS/s sampling rate and a 100 \textmu s acquisition window, requiring a 5 \textmu s pre-trigger window. We set a 5 mV trigger level on the oscilloscope, which is terminated at 50 $\Omega$, to trigger on the rising edge of the pulse. The trigger level was adjusted by eye in order to maximise triggers from real pulses and minimise triggers from electronics noise, and corresponds to approximately 70\% of the raw mean height of a single photon pulse at $V_{\mathrm{bias}}$ = 32 V. Data is stored in a binary file format, saving the waveform data for each trigger, as well as the trigger time.

Each dataset is analysed using a simple pulse finding algorithm, in order to identify all candidate pulses in each 100 \textmu s-long waveform, and not just the pulse which triggered the event. In each 100 \textmu s-long waveform, a moving-average filter with a 80 ns window is applied to smooth out noise fluctuations. A baseline-subtracted waveform is computed using the baseline mean calculated in the first 3.2 \textmu s of the waveform, in the pre-trigger region. We then search for regions of the waveform above a given threshold, which we set to 40\% of the filtered mean amplitude of the single photon pulse for that voltage. We record the segment ``start'' time at which the signal first crosses the threshold, and the segment ``end'' time at which the signal drops below the threshold. This corresponds to the number of consecutive samples where the threshold condition was satisfied, known as the time-over-threshold (ToT). We set a lower threshold in the pulse finding algorithm to ensure that we are sensitive to pulses with amplitudes less than the single photon pulse, which could occur from correlated delayed avalanches occurring within the SiPM recharge time (discussed further in Section~\ref{sub:CDA}). 

\begin{figure}[htpb]
\hspace*{-1cm}  
\centering
\includegraphics[width=1.1\textwidth]{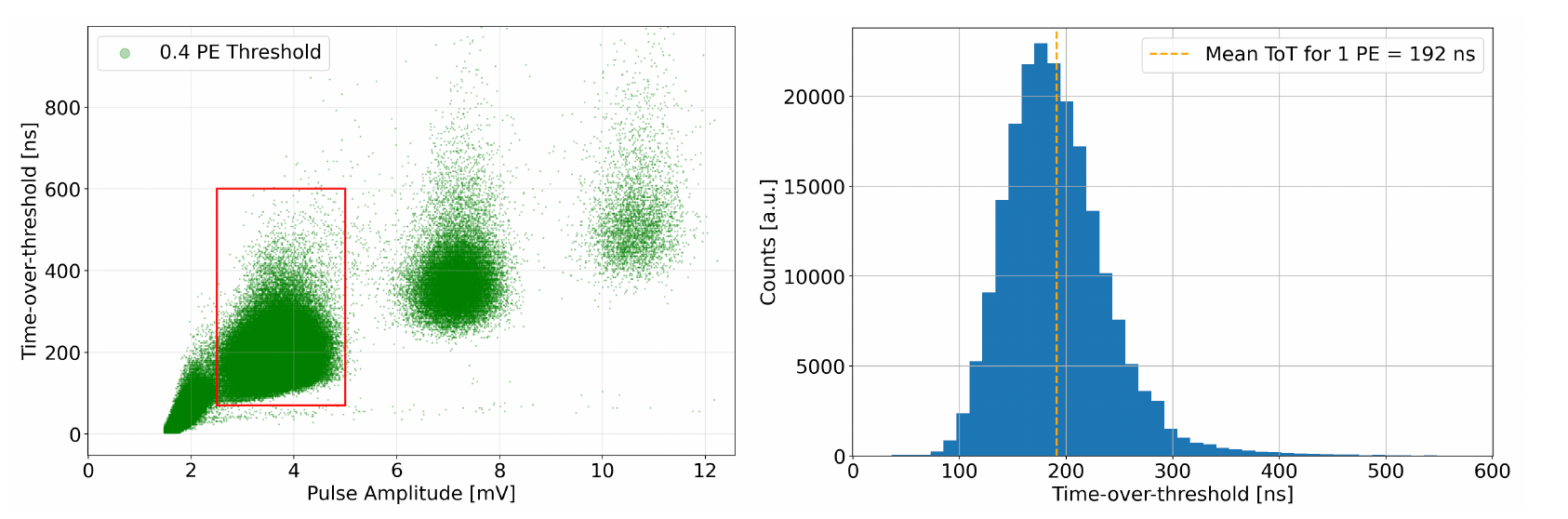}
\caption{Left: Distribution of the time-over-threshold (ToT) versus amplitude for candidate pulses that exceed the pulse finding threshold (40\% of the filtered mean amplitude of the single photon pulse) at $V_{\mathrm{bias}}$ = 32 V. Right: ToT projection of the single photon population illustrated by the red box in the left-hand figure.}
\label{F:PulseFinding}
\qquad
\end{figure}

Figure~\ref{F:PulseFinding} (left) shows the ToT versus amplitude for candidate pulses at $V_{\mathrm{bias}}$ = 32 V. There are three distinct populations which correspond to the 1, 2, and 3 photon pulses. The 1 photon population is slightly asymmetric compared to the 2 and 3 photon populations, extending to lower amplitude and ToT; this likely originates from recharge-limited correlated delayed avalanches. In addition, there is a separate low amplitude-low ToT population, which is likely driven both by recharge-limited correlated delayed avalanches, and noise fluctuations occurring on the tail of a preceding pulse. We decide to apply an additional, but relaxed, requirement on the time-over-threshold (ToT) of $\geq$ 40 ns for pulse finding, to reduce contamination from noise fluctuations.

The consequence of requiring a lower pulse finding threshold is that we have less sensitivity to pulses occurring close together in time; if a secondary pulse occurs before the signal has dropped below the threshold condition from a primary pulse, the pulse finder will not have the sensitivity to resolve the secondary pulse from the primary. Figure~\ref{F:PulseFinding} (right) shows the ToT projection of the single photon population illustrated by the red box in the left-hand figure. The mean ToT for the single photon pulse is 192 ns, which means that on average, secondary pulses that occur within 192 ns of a primary pulse will not be resolved as a separate pulse. The mean ToT value for the single photon population is consistent to within 20 ns across all three $V_{\mathrm{bias}}$ values we operated at. For cases where a secondary pulse occurs below the signal drops below threshold, the pulse finder will identify the pulse with the higher amplitude, which predominately originates from the primary pulse. This can introduce a small bias in the characterisation of the correlated delayed avalanche probability, characterised in Section~\ref{sub:CDA}. 

\begin{figure}[htpb]
\centering
\includegraphics[width=.8\textwidth]{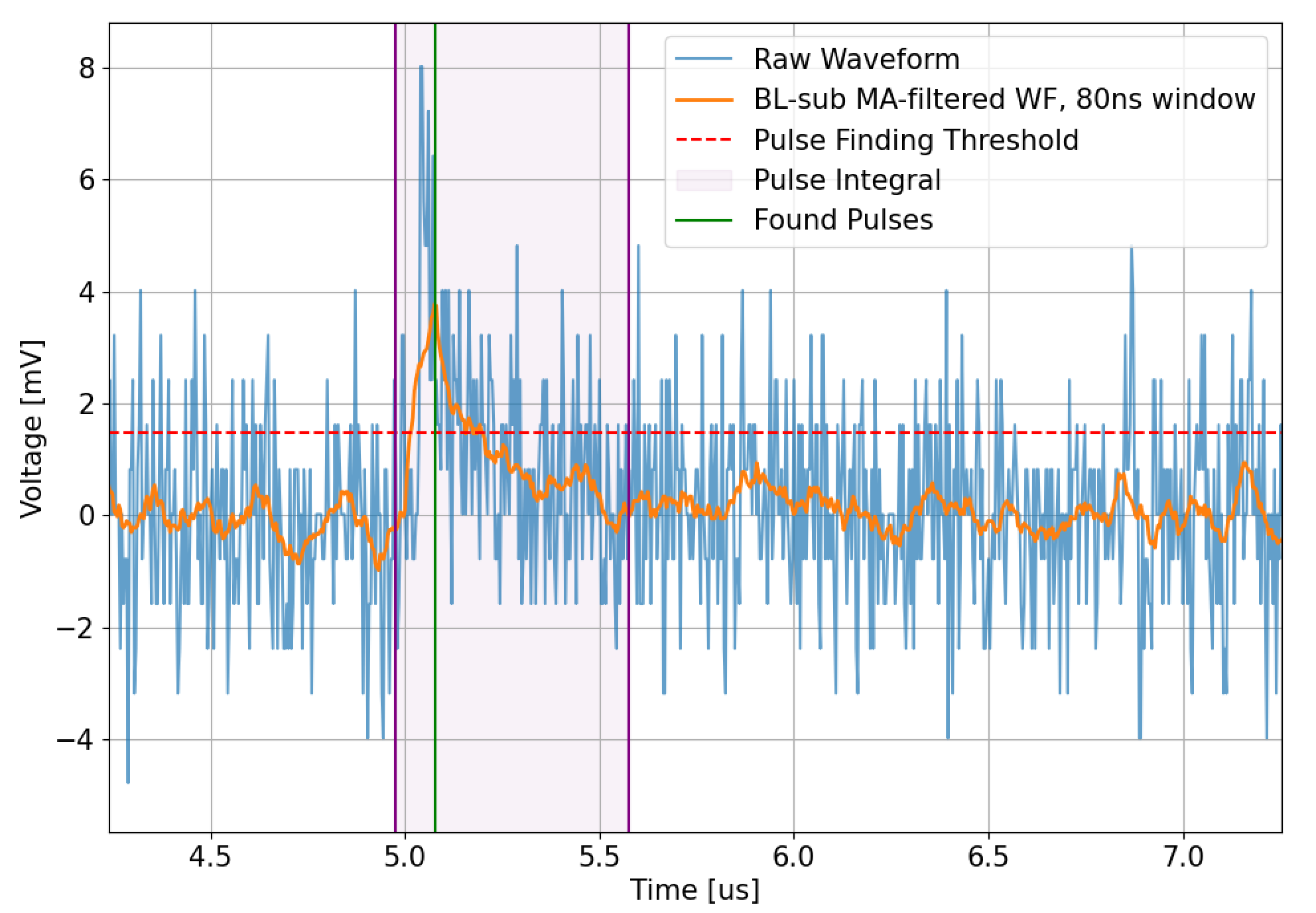}
\caption{Example waveform (WF) acquired at $V_{\mathrm{bias}}$ = 32 V, corresponding to a single dark count rate signal, which is equivalent to the single photon signal. Raw WF (blue), baseline-subtracted (BL-sub) and moving-average (MA) filtered WF (orange), pulse finding threshold (red), and identified pulse time (green) are shown. The shaded purple region corresponds to the region [$t_{\mathrm{p}}$-100 ns : $t_{\mathrm{p}}$+500 ns] where the pulse integral is computed to measure the charge.}
\label{F:Example1PEWF}
\qquad
\end{figure}

Figure~\ref{F:Example1PEWF} shows an example waveform acquired at $V_{\mathrm{bias}}$ = 32 V, corresponding to a single dark count rate signal, which is equivalent to the single photon signal. The raw waveform (WF), the baseline-subtracted (BL-sub) and moving-average filtered waveform (MA-filtered WF), the pulse finding threshold, and the identified pulse time (green) are shown. The pulse time, $t_{\mathrm{p}}$, is defined as the time corresponding to the maximum amplitude. We also calculate a pulse area (proportional to charge) by integrating the waveform between [$t_{\mathrm{p}}-100$ ns : $t_{\mathrm{p}}+500$ ns]; this is illustrated by the shaded purple region. In addition to this information, we determine the absolute time difference between the current pulse and the preceding pulse, $dt$, calculated using the time of the pulses within their respective waveforms, the unix timestamp of the waveforms, and the fraction of a second past the unix timestamps of the waveforms. All of this information is then stored in a data file upon which the downstream characterisation analysis described in Section~\ref{sec:Characterisation} is performed.


\section{Characterisation analysis}
\label{sec:Characterisation}


\subsection{Reverse I-V curves}
\label{sub:IV}
We first measured reverse I-V curves using the Keithley 2450 sourcemeter both at room temperature (prior to the cool-down), and after the cool-down to the mixing chamber base temperature, $T_{\mathrm{CSNT}}$ = 9.4 $\pm$ 0.2 mK. The reverse I-V curve is used to characterise the relationship between the current flowing through the SiPM and the applied bias voltage, and can be used to determine its breakdown voltage, defined as the minimum bias voltage above which the SiPM can operate in Geiger mode and detect single photons. Our main motivation for acquiring reverse I-V curves was to ensure that the SiPM had survived the cool-down to base temperature before we attempted to acquire full characterisation data.

\begin{figure}[htbp]
\centering
\hspace*{-1cm}  
\includegraphics[width=1.1\textwidth]{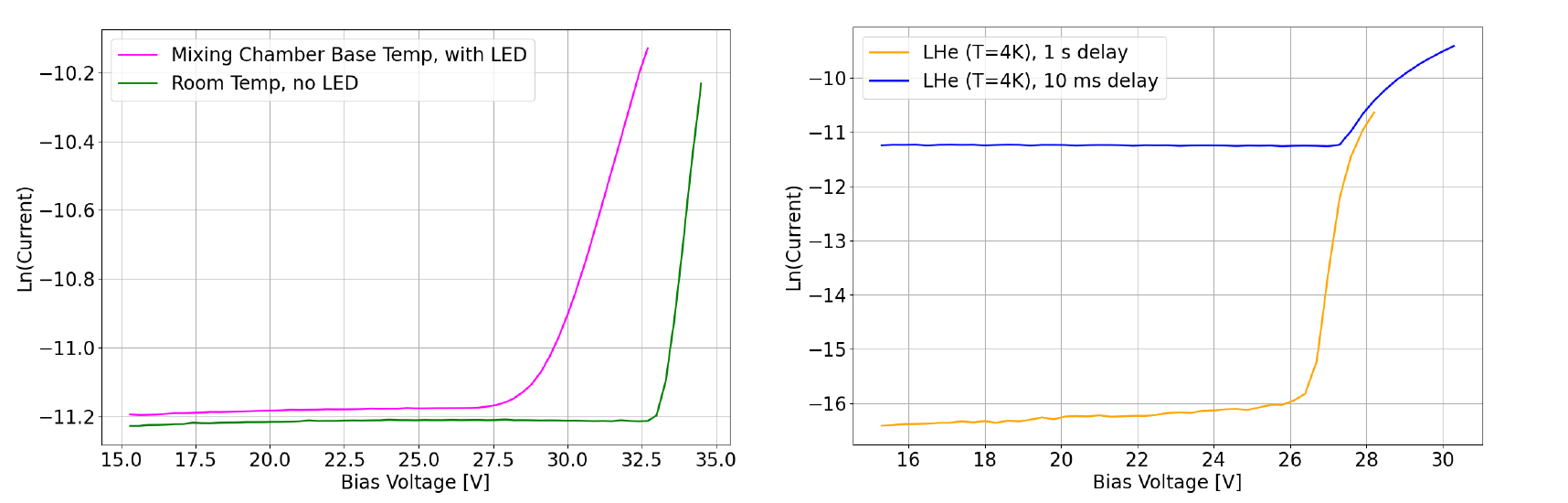}
\caption{Left: Reverse I-V curves taken at room temperature prior to cool-down (green), and at the mixing chamber base temperature post cool-down (magenta). Right: Reverse I-V curves taken whilst the SiPM was submerged in a LHe dewar at $T$ = 4 K, acquired using a delay time of 10 ms (blue), and 1 s (orange) between subsequent voltage steps, demonstrating the impact that the delay time has on both the absolute current value in the linear pre-breakdown regime, and on apparent breakdown voltage.}
\qquad
\label{F:IV_Vbd}
\end{figure}

At room temperature, the dark count rate of the SiPM is large enough such that breakdown can be observed when the SiPM is in dark environment. At cryogenic temperatures however, an external light source is required to trigger breakdown. A signal pulse generator (1 kHz, 999 \textmu s pulse width) and a current-limiting resistor were used to bias the LED whilst the I-V curve was acquired. Both the room temperature and mixing chamber base temperature reverse I-V curves were measured between 0 - 35 V, in 300 mV steps, with a 10 ms time delay between subsequent voltage steps during the sweep. The current limit on the Keithley 2450 was set to 60 \textmu A to avoid damage to the SiPM. Figure~\ref{F:IV_Vbd} (left) shows a comparison of the reverse I-V curve taken at room temperature prior to cool-down, versus at the mixing chamber base temperature post cool-down. We clearly observe the SiPM breakdown both pre and post cool-down, demonstrating that the SiPM survived the cool-down to $T_{\mathrm{CSNT}}$ = 9.4 $\pm$ 0.2 mK. In addition, we observe a clear shift downwards in the breakdown voltage (i.e., the onset of self-sustaining avalanches) between the room temperature measurement and after cool-down, as expected with decreasing temperature. 

We observed that after acquiring the reverse I-V curve, the temperature of the mixing chamber plate increased to $\sim$80 mK. This was measured using a Ruthenium Dioxide (RuO$_{2}$) sensor mounted onto the mixing chamber plate, which measures the temperature accurately down to 20 mK; below 20 mK, the mixing chamber temperature is measured with the CSNT sensor. Given that the SiPM is thermally-coupled to the mixing chamber plate prior to commencing the I-V scans, it is likely that the temperature of the SiPM itself also increased during the sweeps. It is therefore very difficult to determine the breakdown voltage using the reverse I-V curve at a guaranteed temperature of 9.4 mK.


Furthermore, using follow up measurements of the SiPM submersed in a liquid helium (LHe) dewar at $T$ = 4 K, we observed that the reverse I-V curve shape is strongly dependent on the delay time used between subsequent voltage steps during the sweep. Figure~\ref{F:IV_Vbd} (right) shows reverse I-V curves taken in the LHe dewar using a delay time of 10 ms (as was used for the reverse I-V curves taken at the mixing chamber base temperature and at room temperature), versus 1 s. Figure~\ref{F:IV_Vbd}  shows that at $T = 4$ K, there is a clear difference between the reverse I-V curves using delay times of 10 ms and 1 s, both in the absolute current in the linear pre-breakdown regime, and in the apparent breakdown voltage. 

We initially chose a 10 ms delay time between voltage steps as we did not want to inject unnecessary heat into the dilution refrigerator, however, we did not anticipate that there would be such an impact in the reverse I-V curve shape. This is likely driven by an increase in the SiPM quenching resistance at cryogenic temperatures, which lengthens the microcell recharge and thus requires a longer delay time during the sweep in order to get an accurate current measurement. We therefore cannot use the reverse I-V curves to get an accurate measurement of the breakdown voltage, which we also cannot guarantee to be the value of the breakdown at $T_{\mathrm{CSNT}}$ = 9.4 $\pm$ 0.2 mK due to the observed heating during the sweep. As such, we estimate the breakdown voltage using an alternative method, described in Section~\ref{sub:Vbd}.

\subsection{Single photon response and gain}
\label{sub:gain}

We first characterise the single photoelectron (SPE) response of the SiPM. To do this, we plot two histograms corresponding to the pulse amplitude and pulse area for each dataset. As an additional data cleaning cut for the SPE analysis, we only consider pulses where the time difference between the pulse in question and the preceding pulse is $\geq 2$ \textmu s. This is to reject pulses with amplitudes less than the single photon pulse, that can occur from correlated noise or from noise fluctuations on the tails of the previous pulse. 

\begin{figure}[htbp]
\centering
\hspace*{-1cm}  
\includegraphics[width=1.1\textwidth]{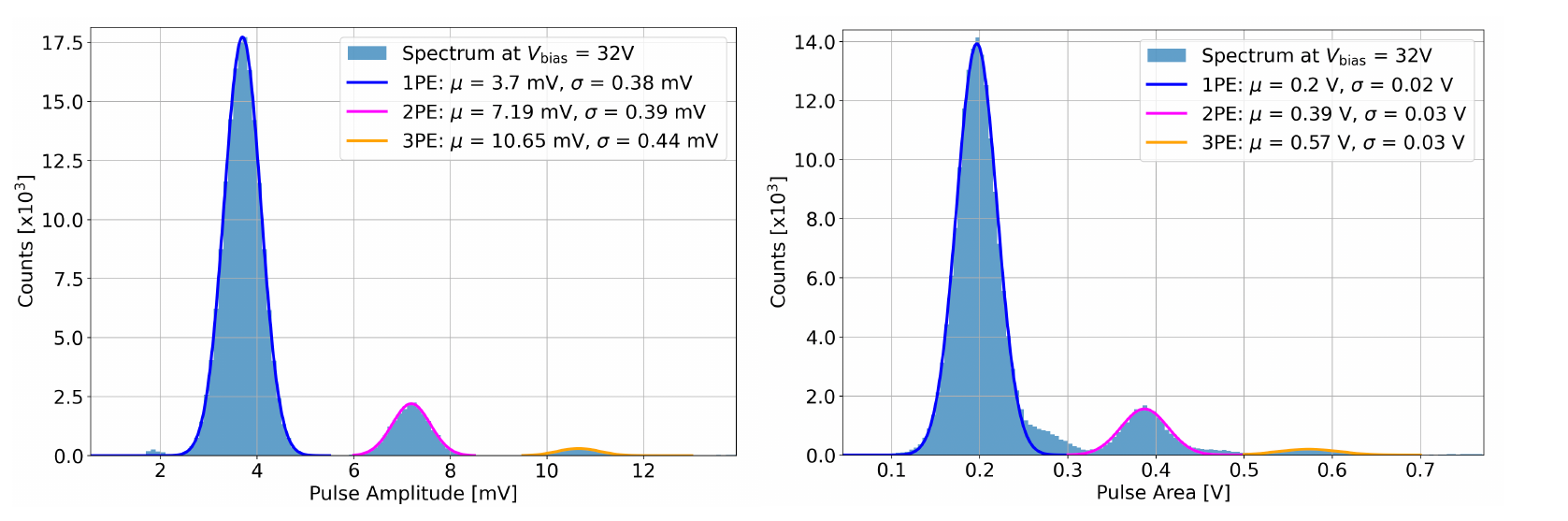}
\caption{Pulse amplitude (left) and area (right) distributions of the NUV-HD-cryo SiPM operated at $V_{\mathrm{bias}}$ = 32 V. Gaussian functions are fit to the first 3 peaks in each histogram. The first peak in each histogram corresponds to the 1 PE peak.}
\label{F:ChargeAmplitudeSpectra_32V}
\end{figure}

Figure~\ref{F:ChargeAmplitudeSpectra_32V} shows the pulse amplitude (left) and pulse area (right) distributions of the SiPM whilst operated at $V_{\mathrm{bias}}$ = 32 V. In both histograms, there are distinct peaks for successive numbers of PE. The first peak is the 1 PE signal, i.e., observed output pulse following the detection of a single photon. Due to intrinsic noise, a SiPM still generates output pulses in the absence of light. The 1 PE peak we observe in our distributions is due to a combination of thermally-induced dark count noise, and noise from correlated delayed avalanches occurring at times later than the SiPM recharge time. These noise sources are discussed in detail in Sections~\ref{sub:DCR} and~\ref{sub:CDA} respectively. The successive peaks corresponding to $>$1 PE we observe in our distributions are a result of correlated noise from direct crosstalk, which manifest as superpositions of the 1 PE waveform. This noise source is discussed further in Section~\ref{sub:DiCT}. 

The mean SPE response is extracted from a gaussian fit to the first peak in each spectrum. The SPE response at $V_{\mathrm{bias}}$ = 32 V is calculated to be $\mu_{\mathrm{SPE,amp}}$ = 3.65 mV $\pm$ 0.42 mV and $\mu_{\mathrm{SPE,area}}$ = 0.15 V $\pm$ 0.02 V in amplitude and area respectively. The mean SPE amplitude is smaller than the raw mean amplitude due to the moving-average filter which is applied to the data prior to running the pulse finding algorithm which reduces the height of the pulses; this can be seen in Figure~\ref{F:Example1PEWF}.

The separation between each pair of consecutive peaks in Figure~\ref{F:ChargeAmplitudeSpectra_32V} (right) is constant and is proportional to the gain, $G$, defined as the amount of charge generated in an avalanche in response to each detected photon. Each microcell in the SiPM produces a highly quantised amount of charge when a photon is absorbed in the active volume, and this response is extremely uniform across all microcells. This unique property of SiPMs results in excellent charge resolution, and allows for a very precise measurement of the gain. The gain is expected to increase proportionally with higher $\Delta V$.

To determine the separation, $S$, we use the results of the gaussian fits to the remaining peaks in the pulse area spectrum. We fit out the separation between peaks as a parameter in a linear polynomial fit to the fitted means of the 1, 2, and 3 PE peaks. The separation is calculated to be $S_{\mathrm{area}}$ = 0.149 $\pm$ 0.0003 V at $V_{\mathrm{bias}}$ = 32 V. $S_{\mathrm{area}}$ is multiplied by the oscilloscope sampling period to convert the separation into units of charge, which is then divided by the electron charge to calculate the gain:

\begin{equation}
G = \frac{S_{\mathrm{area}}[V]\cdot t_{\mathrm{s}}[s]}{R_{\mathrm{scope}}[\Omega]\cdot G_{\mathrm{board}}\cdot q_{\mathrm{e}}[C]},
\qquad
\end{equation}
where $R_{\mathrm{scope}} = 50\ \Omega$ is the input resistance of the oscilloscope, $G_{\mathrm{board}} = 130$ V/V is the gain of the custom amplifier board from the design specification, $q_{\mathrm{e}} = 1.6\times10^{-19}$ C is the electron charge of an electron, and $t_{\mathrm{s}}$ = 4 ns is the sampling period. 

\begin{figure}[htpb]
\centering
\includegraphics[width=0.8\textwidth]{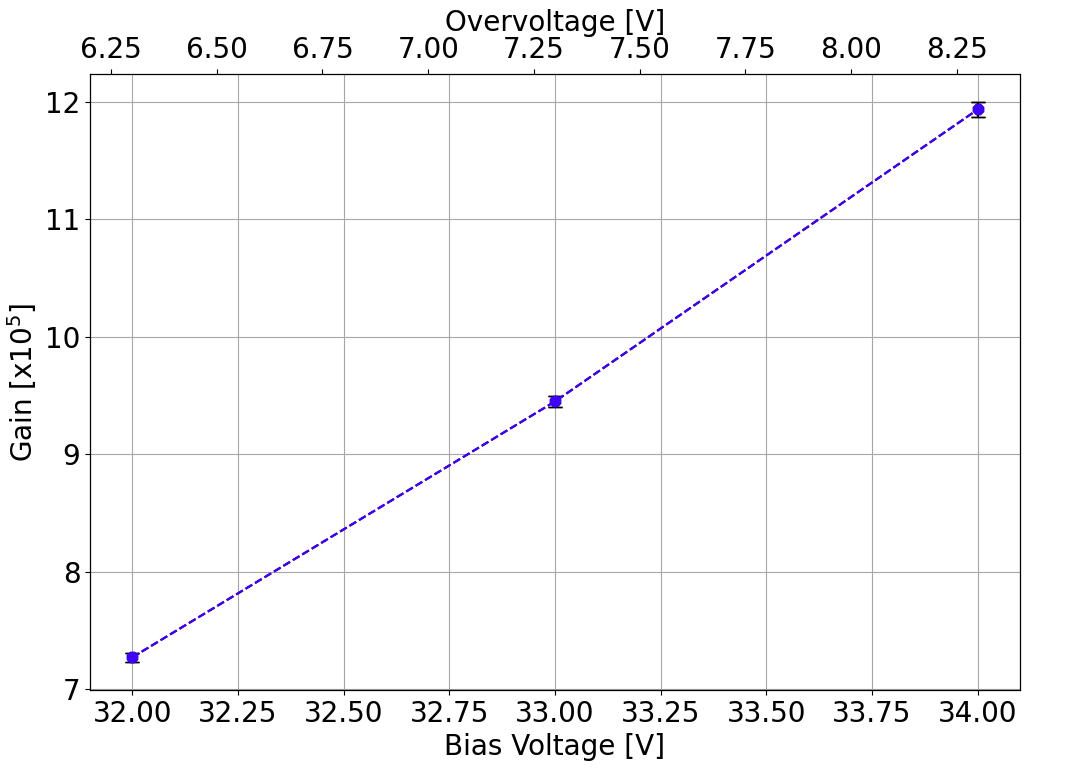}
\caption{Gain of NUV-HD-cryo SiPM as a function of $V_{\mathrm{bias}}$. Statistical error bars are shown, however are not necessarily visible as the errors are small. Blue dashed line is interpolated between the data points.}
\label{F:Gain_versus_V}
\end{figure}

Figure~\ref{F:Gain_versus_V} shows the measured gain as a function of $V_{\mathrm{bias}}$ ($\Delta V$). As expected, we observe a proportional increase in the gain with $\Delta V$. The measured gain ranges between $7\times10^{5}$ and $1.2\times10^{6}$ for 32 V $<V_{\mathrm{bias}}<$ 34 V. At 77 K, the gain is reported to be approximately $G=3.2\times10^{6}$ for the NUV-HD-cryo technology~\cite{acerbi2023nuv}, which is a factor of 3.7 higher compared to our value of $G=8.6\times 10^{5}$ for the same overvoltage value (7 V) based on an interpolation between our data points\footnote{The overvoltage is defined in Section~\ref{sub:Vbd}.}. This difference could be driven by a ``freeze-out'' of free carriers at ultra-low temperatures, a mechanism which has previously been reported at LHe temperatures~\cite{achenbach2018gain}, however we do not explore this further in this manuscript.


\subsection{Breakdown voltage}
\label{sub:Vbd}
To determine the breakdown voltage ($V_{\mathrm{bd}}$) of the SiPM at $T_{\mathrm{CSNT}}$ = 9.4 $\pm$ 0.2 mK, we linearly extrapolate the mean SPE amplitude $\mu_{\mathrm{SPE,amp}}$ down to zero. Since the pulse amplitude is proportional to the prompt avalanche current, it is sensitive to the onset of self-sustaining avalanches and therefore can be used to estimate the breakdown voltage.

We first identify candidate 1 PE signals using the pulse finder algorithm described in Section~\ref{subsec:DAQ}. Similarly to the SPE analysis described in the previous section, to ensure that we obtain a clean sample of 1 PE signals, i.e., excluding recharge-limited pulses with amplitudes $<$ 1 PE or noise fluctuations on the tail, we require that the time between the pulse in question and the previous pulse $\geq 2$ \textmu s. We then search for the maximum amplitude of the pulse in a second filtered version of the waveform, which uses a much shorter MA window of 8 ns (2 samples), within the ToT region identified by the pulse finder.

We do not use the 80 ns MA filtered pulse amplitude for this analysis since the filter reduces the 1 PE pulse amplitude; a longer MA window is intentionally chosen for the pulse finder to more effectively filter out baseline noise, however this does result in a decrease in the pulse height. Using this amplitude in the extrapolation would result in an overestimation of $V_{\mathrm{bd}}$. We therefore use the 8 ns MA filtered amplitude, which is a much closer proxy to the actual height of the 1 PE amplitude. We also repeat the analysis using the raw waveform amplitude, however this results in a large uncertainty on the estimated value of $V_{\mathrm{bd}}$ due to the presence of additional noise in the raw waveforms. We instead use this result as a cross-check with the result obtained from the 8 ns MA filtered amplitude.

\begin{figure}[htpb]
\centering
\hspace*{-1cm}   
\includegraphics[width=1.\textwidth]{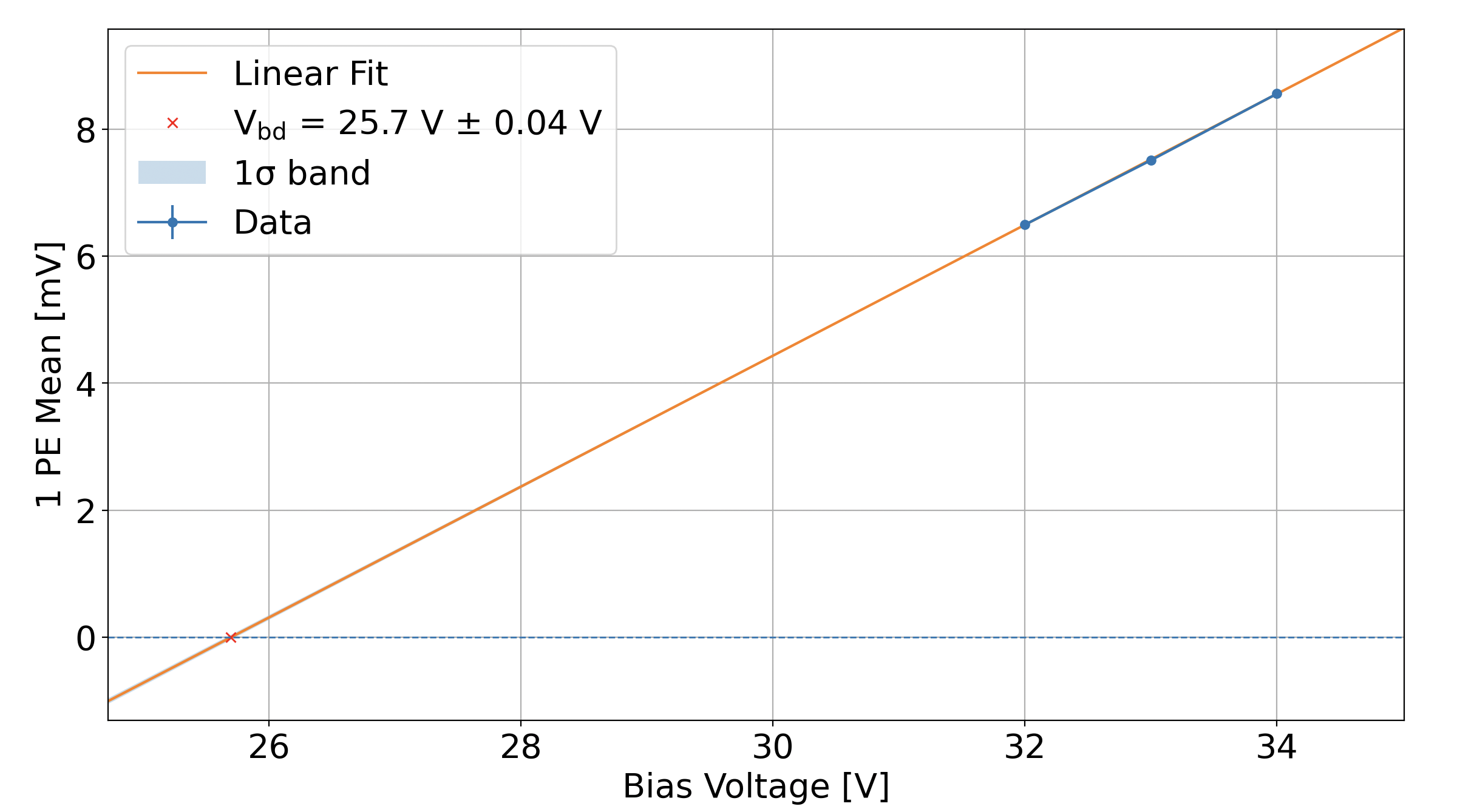}
\caption{Mean SPE amplitude $\mu_{\mathrm{SPE,amp}}$ as a function of $V_{\mathrm{bias}}$, using the maximum amplitude obtained from the 8 ns MA filtered waveform. Linear fit and uncertainty band are also shown. The breakdown voltage extracted from extrapolating $\mu_{\mathrm{SPE,amp}}$ down to zero is $V_{\mathrm{bd}}$ = 25.70 ± 0.04 V.}
\label{F:Vbd_from_1PE_MA}
\qquad
\end{figure}

Similarly to the SPE characterisation, we fit a gaussian to the 1 PE peak amplitude distribution in order to determine $\mu_{\mathrm{SPE,amp}}$ at each bias voltage. We fit these data points with a linear function and extrapolate down to the value of $V_\mathrm{bias}$ at which $\mu_{\mathrm{SPE,amp}}$ goes to zero. Figure~\ref{F:Vbd_from_1PE_MA} shows $\mu_{\mathrm{SPE,amp}}$ (using the 8 ns MA filtered amplitude) as a function of $V_\mathrm{bias}$, including the linear fit and uncertainty band; we determine a breakdown voltage of $V_{\mathrm{bd}}$ = 25.70 ± 0.04 V with this method. Repeating this analysis with the raw waveform amplitude yields a breakdown voltage of $V_{\mathrm{bd}}$ = 25.47 ± 0.58 V. Since our two measurements are consistent within uncertainties, we report the 8 ns filtered MA amplitude result. From hereafter, we report all characterisation measurements both in terms of the bias voltage and in the overvoltage, defined as $\Delta V = V_{\mathrm{bias}} - V_{\mathrm{bd}}$.

The breakdown voltage of the SiPM is expected to decrease with lower temperatures. At 77 K, the breakdown voltage of the NUV-HD-cryo technology is expected to be $V_{\mathrm{bd}}$ = 27.1 V~\cite{NUV-HD-cryo}. As the selected photon sensors for the upcoming DarkSide-20k dark matter experiment, the collaboration has tested 359040 NUV-HD-cryo SiPMs at 77 K. The DarkSide-20k collaboration has determined the average breakdown voltage at 77 K to be $V_{\mathrm{bd}}$ = 27.19 $\pm$ 0.05 V~\cite{QAQC}, which is consistent with the expected value reported in~\cite{NUV-HD-cryo}. The single device we are testing originates from an engineering run used for pre-production assembly, and is of the same type as the DarkSide-20k SiPMs. Therefore, our measurement of $V_{\mathrm{bd}}$ = 25.70 $\pm$ 0.04 V at $T_{\mathrm{CSNT}}$ = 9.4 $\pm$ 0.2 mK is consistent with the expected behaviour of $V_{\mathrm{bd}}$ to decrease with lower temperatures for these devices.

\subsection{Dark count rate}
\label{sub:DCR}
SiPMs can exhibit both uncorrelated and correlated noise contributions. The dark count rate (DCR) is the main source of uncorrelated noise. Each dark count is initiated by a thermally-generated charge carrier, which triggers an avalanche in the active region through the same process as a photon-generated charge carrier; thus DCR events are indistinguishable from single photon signals. The DCR of the SiPM is expected to increase with $\Delta V$~\cite{NUV-HD-cryo}, since a higher $\Delta V$ increases the probability of an avalanche being triggered. The DCR is also expected to increase as a function of temperature. At higher temperatures, the DCR is dominated by field-enhanced thermal generation of charge carriers. However, below a certain temperature, the reduction of the DCR reaches a plateau, where the DCR becomes dominated by tunnelling effects that are weakly temperature dependent. 

We characterise the DCR of the NUV-HD-cryo SiPM using the unshadowing method, based on using the time distribution between two consecutive pulses described in \cite{UnshadowingMethod}. Given the incredibly low DCR of this technology, we need to consider the time differences of pulses across separate waveforms in addition to time differences within waveforms. Assuming a DCR of 5 mHz/mm$^{2}$ at 77 K~\cite{NUV-HD-cryo}, this corresponds to a DCR of 0.5 Hz for a single SiPM. If we were to acquire 600,000 randomly-triggered waveforms, we would expect only 30 dark counts. Thus, to only consider time differences within a waveform would require extremely long waveforms which is not feasible from a data acquisition perspective. We make the assumption that we lose a negligible number of pulses between waveforms as a result of acquiring the data using the Tektronix FastFrame Memory scheme~\cite{FastFrame} described earlier in Section~\ref{subsec:DAQ}, which has a maximum dead time of 2.5\ \textmu s per waveform segment.

To select primary pulses for this analysis, we make two requirements. First, we require $\mu_{\mathrm{SPE,amp}} - 3\sigma_{\mathrm{SPE,amp}} \leq  A_{p_{i}} \leq \mu_{\mathrm{SPE,amp}} + 3\sigma_{\mathrm{SPE,amp}}$, where $A_{p_{i}}$ is the amplitude of the primary pulse, and $\mu_{\mathrm{SPE,amp}}$ and $\sigma_{\mathrm{SPE,amp}}$ are the SPE amplitude mean and standard deviation respectively. Secondly, we require that the time difference between the primary pulse and the previous pulse is $\geq$ 1 ms. This cut is made to reduce the probability of the secondary pulse not being associated with the primary pulse to be negligible, whilst only removing a very small fraction ($\sim$0.05\%\footnote{Based on the time between subsequent events following an exponential distribution for a random Poisson process, and assuming a rate of 5 mHz/mm$^{2} \equiv 0.5$ Hz.}) of DCR events. 

\begin{figure}[htpb]
\centering
\hspace*{-1cm}   
\includegraphics[width=1.1\textwidth]{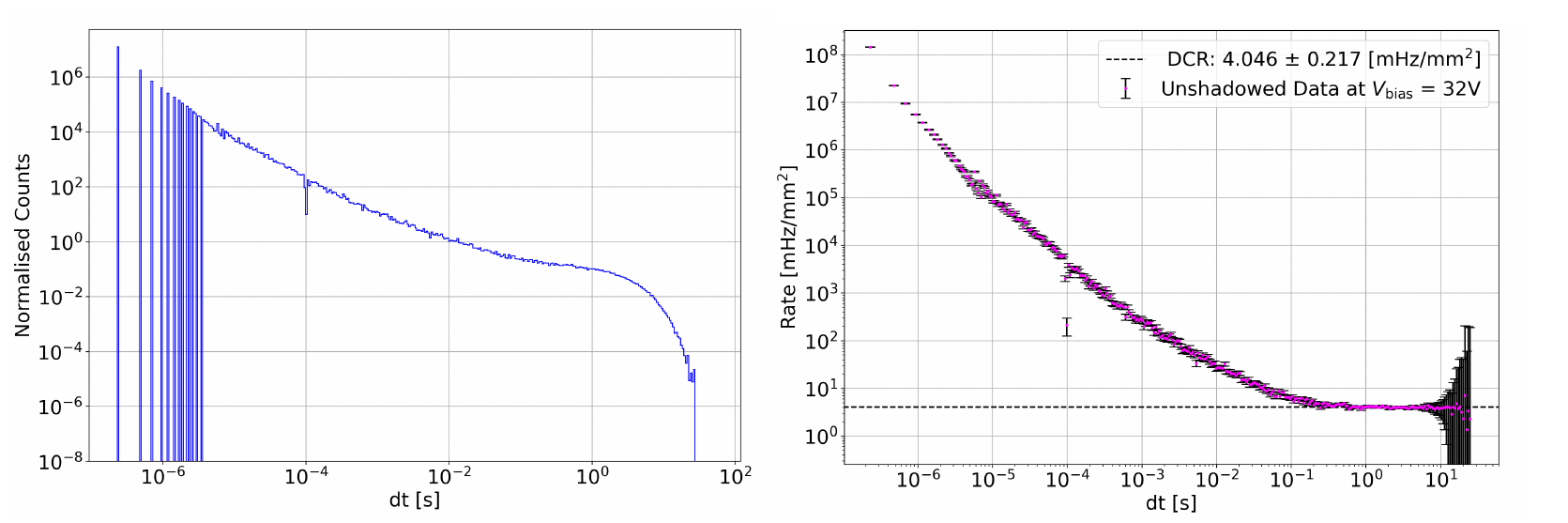}
\caption{Left: Measured time distribution ($dt$) of consecutive pulses at $V_{\mathrm{bias}}$ = 32 V ($\Delta V$ = 6.3 V). Right: Unshadowed time distribution, based on $dt$ distribution in left-hand plot. Horizontal dashed line indicates calculated DCR of 4.046 $\pm$ 0.217 mHz/mm$^{2}$.}
\label{F:DCR_32V}
\qquad
\end{figure}

Figure~\ref{F:DCR_32V} (left) shows the measured time distribution of consecutive pulses whilst the SiPM is operated at $V_{\mathrm{bias}}$ = 32 V ($\Delta V$ = 6.3 V). The dip in the $dt$ distribution at close to the waveform acquisition window, 100~\textmu s, can be attributed to a loss in the pulse finder efficiency for pulses that occur near the end or beginning of waveform that may be partially cut-off, as well as from the dead time between waveforms itself, showing that the loss of pulses between waveforms is not necessarily negligible. We also observe that if we apply a stricter (larger) $dt$ cut between the primary pulse and the preceding pulse, the effect of the dip is slightly enhanced.

\begin{figure}[htpb]
\centering
\includegraphics[width=0.8\textwidth]{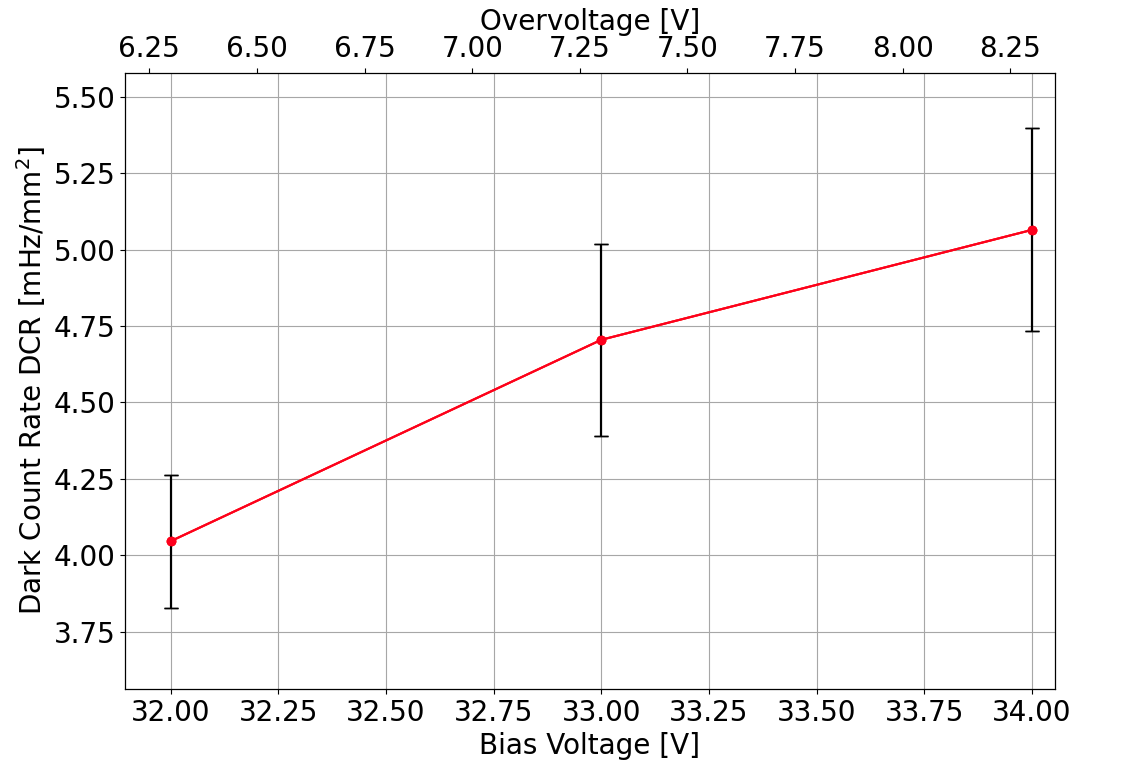}
\caption{Measured DCR as a function of $V_{\mathrm{bias}}$ ($\Delta V$). The DCR increases by 25\% between 6.3 V $< \Delta V <$ 8.3 V. Red line is interpolated between the data points.}
\label{F:DCR_vs V}
\qquad
\end{figure}

Figure~\ref{F:DCR_32V} (right) shows the unshadowed time distribution following the procedure in~\cite{UnshadowingMethod}. The unshadowed distribution is the average number of secondary pulses occurring between the bin limits of $t_{i}$ and $t_{i+1}$ defined in Figure~\ref{F:DCR_32V} (left), normalised per unit time to determine the pulse rate. The pulse rate distribution contains two contributions: the correlated delayed pulse rate, and the DCR. At large times, the rate of correlated delayed pulses is expected to disappear, leaving only the DCR. To extract the DCR, we compute the weighted mean of the asymptotic rate, between 1 s $ <dt<$ 8 s. We calculate the DCR to be 4.046 $\pm$ 0.217 mHz/mm$^{2}$ at $V_{\mathrm{bias}}$ = 32 V ($\Delta V$ = 6.3 V), the value of which is indicated by the horizontal dashed line in Figure~\ref{F:DCR_32V} to guide the eye. Figure \ref{F:DCR_vs V} shows the measured DCR as a function of $V_{\mathrm{bias}}$ ($\Delta V$). As expected, the DCR increases as a function of $\Delta V$, increasing by 25\% for 6.3 V $< \Delta V <$ 8.3 V. 

We compute a cross-check of our DCR measurements from the unshadowed method by calculating $N_{\mathrm{primaries}}/\mathrm{livetime}$ for each voltage, where $N_{\mathrm{primaries}}$ is the number of selected primary pulses, and livetime is defined as the total real acquisition time minus dead time. Assuming a maximum dead time of 2.5 \textmu s per waveform segment, this corresponds to $<$ 0.0002\% of the total acquisition time per dataset, and thus is negligible. With this method, we calculate DCR values of 4.025 mHz/mm$^{2}$, 4.516 mHz/mm$^{2}$, and 4.594 mHz/mm$^{2}$ for  $V_{\mathrm{bias}}$  = 32 V, 33 V, and 34 V respectively. Based on the DCR values calculated from the unshadowed method shown in Figure~\ref{F:DCR_vs V}, we determine that these two independent DCR are consistent to within 0.1$\sigma$, 0.6$\sigma$, and 1.4$\sigma$ for these three voltages respectively.

We note that the DCR values we calculate here are also consistent with the measured DCR value of $<$5 mHz/mm$^{2}$ at 77 K from~\cite{NUV-HD-cryo}, which could suggest that for temperatures below 77 K, the DCR is approximately constant. However, this is based on our measurement of a single device at $T_{\mathrm{CSNT}}$ = 9.4 $\pm$ 0.2 mK; measurements of several devices at more temperature points between 77 K and our base temperature would be required to confirm this.

\subsection{Direct crosstalk}
\label{sub:DiCT}
Optical crosstalk is a source of correlated noise in a SiPM that occurs due to the emission of a small number of photons ($\sim$10$^{-5}$ photons/electron) during an avalanche~\cite{mirzoyan2009light}, which can trigger avalanches in the surrounding microcells. Optical crosstalk can be split into two categories:

\begin{enumerate}
    \item Direct crosstalk (DiCT) occurs when a photon directly generates charge carriers in the active region of a neighbouring microcell, resulting in an almost simultaneous avalanche with respect to the primary avalanche.
    \item Delayed crosstalk (DeCT) occurs when a photon generates charge carriers outside of the active region of a neighbouring microcell. The charge carriers can diffuse to the active region of the microcell, triggering a delayed avalanche.
\end{enumerate}

DiCT manifests as a superposition with the 1 PE (single photon) pulse. For example, if a primary avalanche triggers two secondary crosstalk avalanches, the resulting output pulse will be a 3 PE pulse. DiCT can therefore be a limiting factor for applications requiring accurate photon-counting, since it artificially increases the number of detected photons and degrades the signal-to-noise ratio. The DiCT probability is expected to increase with $\Delta V$, since the probability of an avalanche increases with 
$\Delta V$. In addition, the number of emitted photons is linearly proportional to $\Delta V$, which also contributes to an increase in the DiCT probability. There appears to be a weak dependence on the DiCT probability with temperature for the NUV-HD-cryo SiPM technology, as reported in~\cite{NUV-HD-cryo}.

\begin{figure}[htpb]
\centering
\hspace*{-1cm}  
\includegraphics[width=1.1\textwidth]{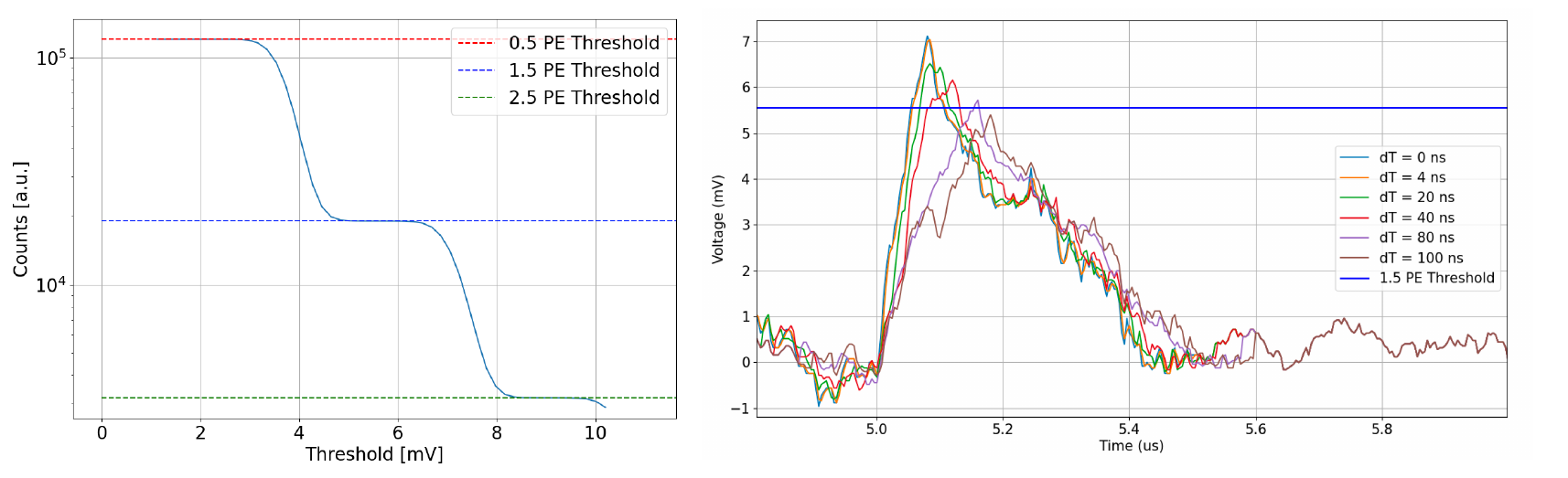}
\caption{Left: Pulse count rate as a function of threshold for $V_{\mathrm{bias}}$ = 32 V ($\Delta V$ = 6.3 V). As the threshold is increased, the count rate decreases in a step-like fashion, leading to a characteristic ``staircase''. Right: Study to explore the effect on pulse reconstruction for delayed avalanches closely following a primary, obtained by selecting a 1 PE waveform $V_{\mathrm{bias}}$ = 32 V ($\Delta V$ = 6.3 V), duplicating and superimposing the waveform onto the original, and shifting it in time to see the time difference at which the reconstructed amplitude drops below the DiCT threshold of 1.5 PE.}
\label{F:DiCT}
\qquad
\end{figure}

To characterise the DiCT of the SiPM, we measure the pulse count rate as a function of threshold within a prompt coincidence window. As the threshold is increased, the count rate decreases in a step-like fashion, leading to a characteristic ``staircase'' where each step corresponds to excluding events with increasing numbers of fired microcells (1 PE, 2 PE, etc). The rate measured at the top step, $R_{\mathrm{0.5PE}}$, is due to all detectable pulses, whilst the rate measured at the second step, $R_{\mathrm{1.5PE}}$, is due to DiCT pulses. The probability of DiCT, $P_{\mathrm{DiCT}}$, can then be defined as:

\begin{equation}
    P_{\mathrm{DiCT}} = \frac{R_{\mathrm{1.5PE}}}{R_{\mathrm{0.5PE}}},
\end{equation}
Figure~\ref{F:DiCT} (left) shows the characteristic staircase obtained at $V_{\mathrm{bias}}$ = 32 V ($\Delta V$ = 6.3 V). By taking the ratio of the blue and red dashed lines, we determine $P_{\mathrm{DiCT}}$ = 15.8\%.

For this analysis, we only consider pulses which triggered the waveform, since we know the pulse trigger time. We first plot the distribution of trigger pulse amplitude times to identify the mean time; we only accept pulses for the DiCT analysis with amplitudes that reconstruct within 20 ns of the mean trigger pulse time. This requirement is made to reduce the bias on the DiCT measurement from DeCT pulses occurring very shortly after the primary pulse, in which case the MA filter and pulse finder will merge the pulses. We performed a test in which we: obtained a single 1 PE waveform at $V_{\mathrm{bias}}$ = 32 V; duplicated the waveform about the pulse and superimposed it on the original pulse (at $t_{\mathrm{p}}$ = 0); shifted it in time to study the change in the pulse shape. This is illustrated in Figure~\ref{F:DiCT} (right), which shows that for this particular waveform, two pulses separated by $\leq$ 80 ns will merge and reconstruct with an amplitude surpassing the 1.5 PE threshold. 

\begin{figure}[htpb]
\centering
\includegraphics[width=0.8\textwidth]{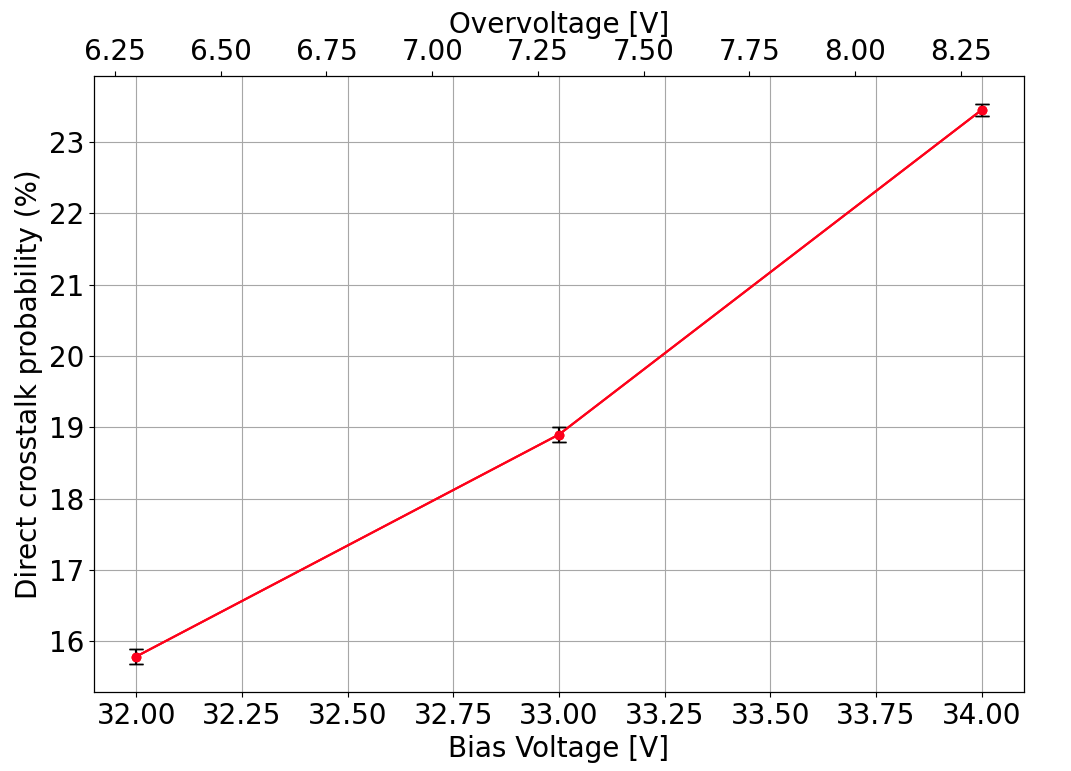}
\caption{$P_{\mathrm{DiCT}}$ as a function of $V_{\mathrm{bias}}$ ($\Delta V$). $P_{\mathrm{DiCT}}$ increases by 49\% between 6.3 V $< \Delta V <$ 8.3 V. Solid red line is interpolated between the data points. Statistical error bars are shown, however are not necessarily visible as the errors are small.}
\label{F:DiCT_vs_V}
\qquad
\end{figure}

Figure~\ref{F:DiCT_vs_V} shows $P_{\mathrm{DiCT}}$ as a function of $\Delta V$. As expected, $P_{\mathrm{DiCT}}$ increases as a function of $\Delta V$, increasing by 49\% across 6.3 V $< \Delta V <$ 8.3 V. At 77 K, the DiCT probability is reported to be 13\% at $\Delta V$ = 5 V for the NUV-HD-cryo technology~\cite{NUV-HD-cryo}, compared to our measurement of 15.8\% at $\Delta V= $ 6.3 V. Our measurement is approximately 20\% higher, however we are also operating at $\Delta V$ about 1 V higher; it is plausible that at $\Delta V= $ 5 V, our measurement would be consistent with the value reported in~\cite{NUV-HD-cryo}. Thus, based on the single device we have tested, there is no clear temperature dependence on $P_{\mathrm{DiCT}}$ for the NUV-HD-cryo technology.



\subsection{Correlated delayed avalanches}
\label{sub:CDA}
Correlated delayed avalanches (CDAs) are a type of correlated noise in SiPMs, comprised of DeCT and afterpulses (APs). DeCT events occur within very short timescales, typically $\mathcal{O}(\mathrm{10\ ns})$ after the primary avalanche, whereas AP events tend to occur on longer timescales, $\mathcal{O}$(ns - \textmu s).

Afterpulsing is a process in which charge carriers generated in an avalanche can become trapped within defects of the silicon structure. Trapped charge carriers are then subsequently released, which can trigger avalanches in the same microcell after the primary avalanche has occurred. A single primary avalanche can result in multiple APs as a result of numerous trapped charge carriers being released at different timescales, or due to an AP triggering a second AP. If the timescale of the AP is short compared to the characteristic recharge time of the SiPM, the gain of the AP-induced avalanche is smaller than the primary avalanche, resulting in a pulse amplitude less than the 1 PE amplitude. If the timescale of the AP is long compared to the characteristic recharge time of the SiPM, APs are indistinguishable from 1 PE signals induced either thermally (from DCR) or optically (from a photon). 

Similarly to DiCT, APs artificially increase the number of detected photons. APs can also distort integration measurements for high photon fluxes, effectively serving as an additional gain factor for the measured signal. The AP probability is expected to increase with $\Delta V$~\cite{NUV-HD-cryo}, since a higher $\Delta V$ results in a larger initial avalanche charge. This, combined with an increase in the probability of an avalanche, results in a higher AP probability.

\begin{figure}[htpb]
\centering
\hspace*{-1cm} 
\includegraphics[width=1.1\textwidth]{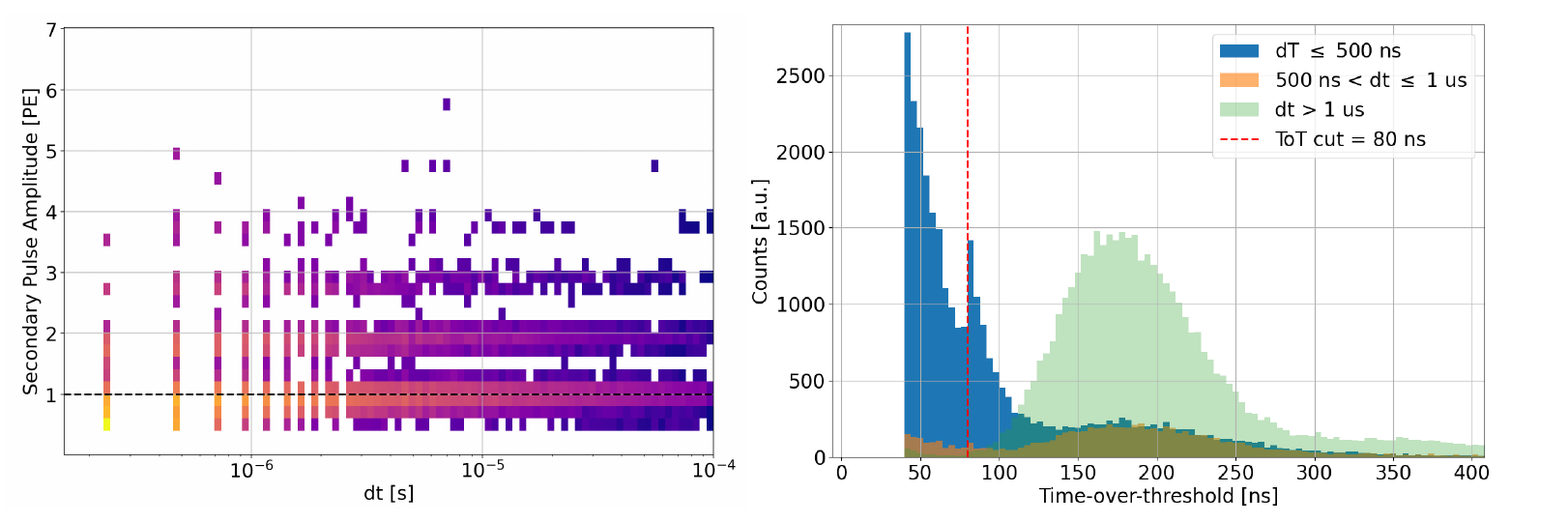}
\caption{Left: 2D distribution of secondary pulse amplitudes versus $dt$ for $V_{\mathrm{bias}}$ = 32 V ($\Delta V$ = 6.3 V), showing recharge-limited APs occurring within $dt <$ 1 \textmu s of a primary pulse. Right: ToT distribution for pulses occurring at $dt \leq$ 500 ns (blue), 500 ns $< dt \leq$ 1 \textmu s (orange), and $dt >$ 1 \textmu s (green) with respect to a primary pulse for $V_{\mathrm{bias}}$ = 32 V ($\Delta V$ = 6.3 V). Red dashed line corresponds to the ToT value above which we accept pulses (with $dt \leq$ 1 \textmu s) for the CDA analysis.}
\label{F:SecondaryPk_ToT}
\qquad
\end{figure}

We determine the mean number of CDAs, $\bar{N}_{\mathrm{CDA}}$, within three different time windows following a primary avalanche: 1 \textmu s, 5~\textmu s, and 10~\textmu s. The mean number of CDAs is calculated on a waveform-by-waveform basis. For each waveform, we count how many subsequent pulses are found within a 1 \textmu s, 5~\textmu s, and 10~\textmu s window after the primary pulse (i.e., the first pulse in the waveform that caused the trigger to occur). We apply no requirement on the amplitude of pulses found $<$1 \textmu s with respect to the primary pulse, since the gain is decreased for APs occurring within the SiPM recharge time. This can be seen in Figure~\ref{F:SecondaryPk_ToT} (left), which shows the 2D distribution of secondary pulse amplitudes versus $dt$ for $V_{\mathrm{bias}}$ = 32 V ($\Delta V$ = 6.3 V). 

However, we do apply a ToT requirement on pulses detected $<$1 \textmu s from the primary pulse to reduce contamination from noise fluctuations occurring on pulse tails. Figure~\ref{F:SecondaryPk_ToT} (right) shows the ToT distribution for pulses occurring at $dt \leq$ 500 ns, 500 ns $< dt \leq$ 1 \textmu s, and $dt >$ 1 \textmu s with respect to the primary pulse for $V_{\mathrm{bias}}$ = 32 V ($\Delta V$ = 6.3 V). There is an increase in identified pulses for $dt$ values less than and on the order of the average length of a single photon pulse ($\sim$500 ns), a population which decreases with larger $dt$. We decide to exclude pulses from the CDA analysis with ToT < 80 ns for $dt \leq$ 1 \textmu s; this aligns with the second peak visible in the blue distribution of Figure~\ref{F:SecondaryPk_ToT}, above which we believe pulses are dominated by real CDAs rather than noise from tail fluctuations.

For pulses detected $\geq$1 \textmu s from the primary pulse, we apply a requirement on the pulse amplitude to be $\geq \mu _{\mathrm{SPE,amp}} - 3\sigma_{\mathrm{SPE,amp}}$, to reduce leakage from noise. We make one additional requirement, that the time difference between the primary pulse of a waveform $p_{i}$ and the last pulse of the preceding waveform $p_{i-1}$ is $\geq$ 1 s. This cut is made to ensure that the primary pulse originates from DCR or from a photon-induced PE, and is not a CDA of a previous pulse. The choice of 1 s is informed from the unshadowed time distribution in Figure~\ref{F:DCR_32V}, which shows the asymptotic rate from pure DCR at time differences $\geq$1 s.

\begin{figure}[htpb]
\centering
\includegraphics[width=0.8\textwidth]{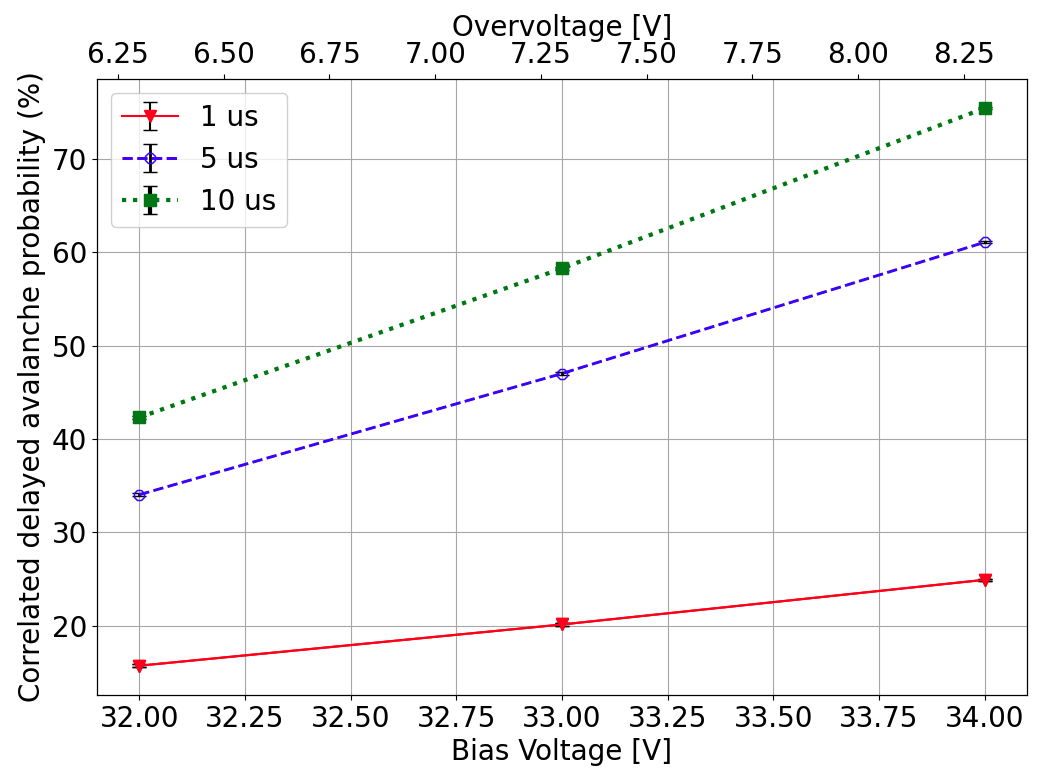}
\caption{Correlated delayed avalanche probability ($P_{\mathrm{CDA}}$) as a function of $\Delta V$ considering three time windows following a primary avalanche: 1 \textmu s (red), 5~\textmu s (blue), and 10~\textmu s (green). $P_{\mathrm{CDA}}$ increases by 59\% (1 \textmu s window), 80\% (5 \textmu s window), and 79\% (10 \textmu s window) between $\Delta V$ = 6.3 V and $\Delta V$ = 8.3 V. Solid red, dashed blue, and dotted green lines are interpolated between the data points. Statistical error bars are shown, however are not necessarily visible as the errors are small.}
\label{F:CDA_vs_V}
\qquad
\end{figure}

Figure~\ref{F:CDA_vs_V} shows the CDA probability, $P_{\mathrm{CDA}}$, as a function of $\Delta V$ within a 1 \textmu s (red), 5~\textmu s (blue), and 10~\textmu s (green) window following a primary avalanche\footnote{The CDA probability is calculated using the mean number of correlated delayed avalanches with the following equation: $P_{\mathrm{CDA}} = 1\ -\ e^{^-\bar{N}_{\mathrm{CDA}}}$.}. The CDA probability is measured to be $P_{\mathrm{CDA}}$ = 15.7\% $\pm$ 0.2\% (1 \textmu s window), 34.0\% $\pm$ 0.3\% (5 \textmu s window), and 42.3\% $\pm$ 0.3\% (10 \textmu s window) at $\Delta V$ = 6.3 V. $P_{\mathrm{CDA}}$ increases by 59\% (1 \textmu s window), 80\% (5 \textmu s window), and 79\% (10 \textmu s window) between $\Delta V$ = 6.3 V and $\Delta V$ = 8.3 V. However, we do note that as our pulse finder is unable to resolve pulses that occur (on average) less than approximately 200 ns (Figure~\ref{F:PulseFinding}), there could be a systematic underestimation of $P_{\mathrm{CDA}}$ in our results. $P_{\mathrm{CDA}}$ increases as a function of $\Delta V$ as expected, driven by the increase in AP probability for higher $\Delta V$. We also observe an increase in $P_{\mathrm{CDA}}$ for longer time windows following a primary avalanche for a given $\Delta V$ value.

\subsubsection{Enhanced afterpulsing at low temperatures}
\label{subsec:APtrains}
At 77 K, the AP probability is reported to be 12\% at $\Delta V$ = 5 V~\cite{NUV-HD-cryo}. We determine a CDA probability of 42\% at $\Delta V$ = 6.3 V assuming a 10 \textmu s integration window, as is typically considered for computing AP probabilities in SiPMs~\cite{acerbi2015nuv,acerbi2019understanding}. We assume that the number of DeCT events contributing to our $P_{\mathrm{CDA}}$ measurement is negligible; we make this assumption based on the timescale of DeCT events, which occur at $dt<$ 100 ns, which we do not have the ability to resolve from the primary avalanche in our data. Therefore, we assume that our $P_{\mathrm{CDA}}$ measurement is dominated by APs.

We observe a substantial increase in the AP probability at $T_{\mathrm{CSNT}}$ = 9.4 $\pm$ 0.2 mK compared to the measured value at 77 K. This behaviour has previously been reported at temperatures below 40 K~\cite{hanski2025performance,zhang2022scintillation}. It has also been observed that below 40 K, the average time difference between a primary avalanche and AP(s) increases. We also see evidence for an increase in the delay times of APs with respect to the primary avalanche. The $dt$ distribution shown in Figure~\ref{F:DCR_32V} shows long tails that extend significantly beyond the typical AP timescale of 10 \textmu s. This is indicative of delayed APs that continue to occur even at much longer timescales. 

The combination of a higher AP probability and an increase in the average time delay of APs can result in long-lasting AP ``trains'', in which APs trigger further APs, causing a self-sustaining cascade of APs that can last for a long time at ultra-low temperatures, up to approximately 1 ms~\cite{hanski2025performance}. This effect is driven by the temperature-dependent trapping and release of charge carriers. During an avalanche, a fraction of carriers can be captured by deep-level traps in the silicon bulk and later released with a characteristic time. As the temperature decreases, the thermal emission rate from these traps is strongly suppressed, leading to much longer de-trapping times. When released, these carriers can retrigger avalanches, producing APs. At sufficiently low temperatures, a single primary avalanche can populate multiple traps with a range of lifetimes, resulting in a sequence of delayed releases and thus a correlated train of APs rather than a single event. This mechanism accounts for both the increased AP probability and the emergence of long AP trains at ultra-low temperature operation.

\begin{figure}[htpb]
\centering
\includegraphics[width=0.9\textwidth]{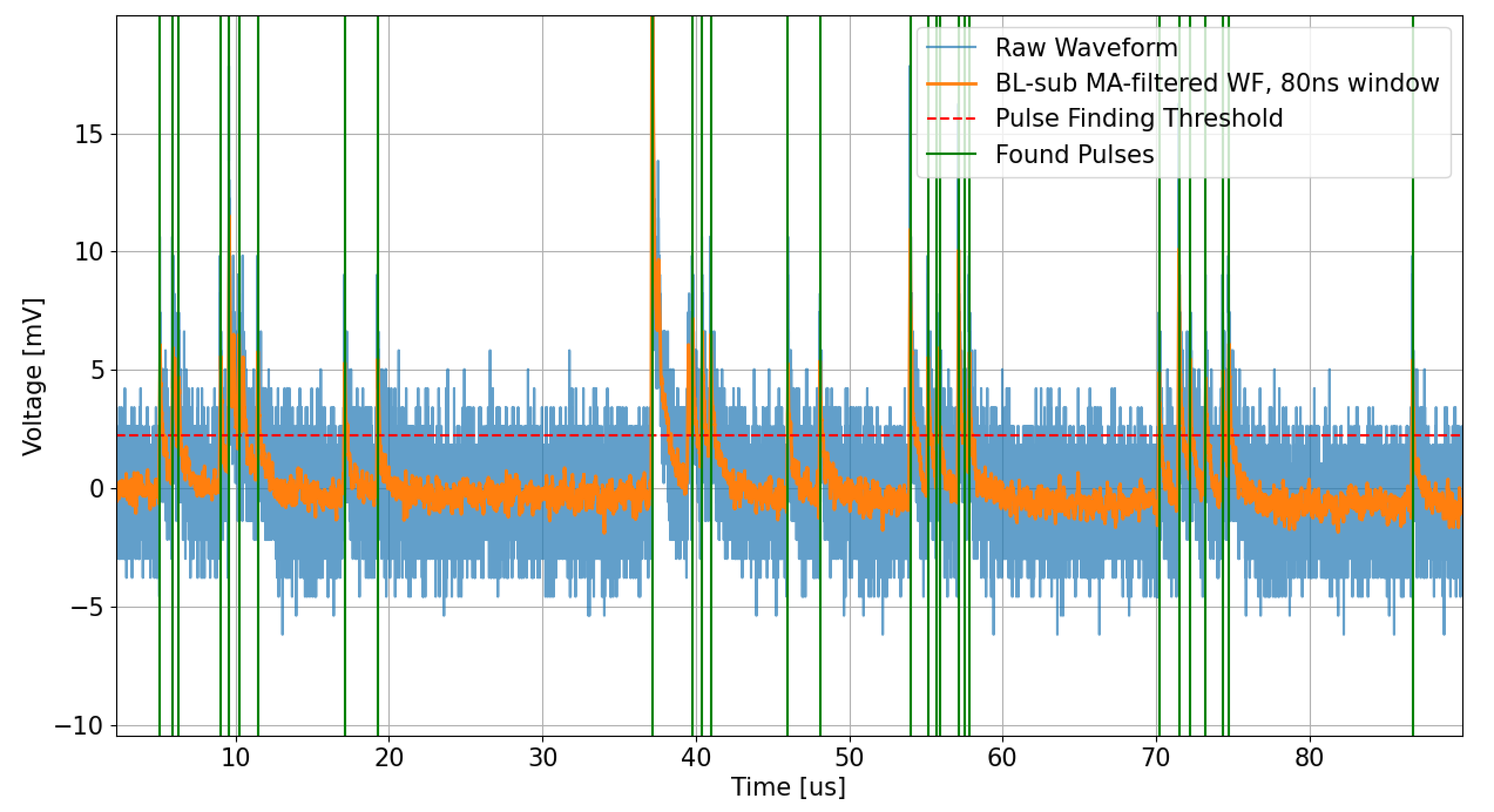}
\caption{Example waveform trace acquired at $V_{\mathrm{bias}} = $ 34 V ($\Delta V$ = 8.3 V), illustrating long-lasting APs consistent with a self-sustaining AP train. Trace shows raw waveform (blue), baseline-subtracted moving-average filtered waveform (orange), pulse finding threshold (dashed red), and identified pulse times (green).}
\label{F:WF_34V_Train}
\qquad
\end{figure}

Figure~\ref{F:WF_34V_Train} shows an example waveform trace acquired at $V_{\mathrm{bias}} = $ 34 V ($\Delta V$ = 8.3 V), consistent with a self-sustaining AP train. The presence of such AP trains clearly limits the photon counting ability of the SiPM, particularly at higher $\Delta V$ values. Whilst delayed APs will also distort integration measurements, Figure~\ref{F:CDA_vs_V} shows that the level of distortion caused by delayed APs can be mitigated by using shorter integration windows and operating at a lower $\Delta V$.

\section{Towards a cryogenic cosmic-ray muon veto system}
\label{sec:cosmicveto}

The main motivation for testing the NUV-HD-cryo SiPM technology in an ultra-low temperature environment is to determine whether it is feasible for SiPMs to be deployed as photon sensors for the cryogenic cosmic-ray muon veto system currently being designed for the QUEST-DMC experiment. As outlined in Section~\ref{sec:Introduction}, the concept design makes use of a scintillator volume surrounding the bolometer box holding the superfluid $^{3}$He target, coupled to a SiPM. High energy cosmic-ray muons are minimum ionising particles which deposit approximately 2 MeV/g/cm$^{2}$ of energy in materials they traverse. If a cosmic-ray muon passes through QUEST-DMC, it will also deposit a fraction of its energy into the scintillator surrounding the experiment, producing a flux of scintillation photons. These can be used as a veto signal, to remove any fiducial volume events from the dark matter search that occur in coincidence with a scintillator signal consistent with a cosmic-ray muon. If the SiPM is directly coupled to the scintillator (i.e., without using fibres to optically couple the physically separated scintillator and SiPM), it will need to be capable of operating at the same temperature as the scintillator, which is thermally coupled to a thermal radiation shield at a temperature of $\mathcal{O}$(10s mK).

The requirements of the cryogenic cosmic-ray muon veto system are set by both its primary function of tagging muon-induced events, and the constraints of operation near the main experimental volume at mK temperatures. The system must provide a sufficiently large response to muon energy deposits to allow a practical threshold to be defined in the presence of background interactions from environmental gammas, targeting a veto efficiency of $\geq$90\% while keeping accidental vetoes - and hence experimental dead time - minimal. A detailed characterisation of the SiPM noise is essential, as it sets the minimum achievable trigger threshold. In particular, correlated noise (DiCT, DeCT, APs) can increase the apparent charge of background signals and raise the accidental trigger rate, limiting how low the threshold can be set. In addition, the SiPM must operate stably at mK temperatures, meeting constraints on power dissipation and thermal load of the dilution fridge, and remaining compatible with the low-noise, low-background environment of the experiment.

In addition to the main dataset used for the characterisation analysis, we also acquired a second, shorter dataset where a small volume of scintillator was coupled to the SiPM. We acquired 2500 waveforms per voltage under the same trigger conditions as before, with a shorter 10 \textmu s acquisition window (1 \textmu s pre-trigger, 9 \textmu s post-trigger). The plastic scintillator used for this study was an early prototype for the SoLid experiment~\cite{abreu2018optimisation}, and is reported to have a scintillation efficiency of approximately 10000 photons/MeV e$^{-}$ at room temperature. This was a convenient test sample for the purpose of these initial proof-of-concept measurements.

\begin{figure}[htpb]
\centering
\hspace*{-1cm} 
\includegraphics[width=1.1\textwidth]{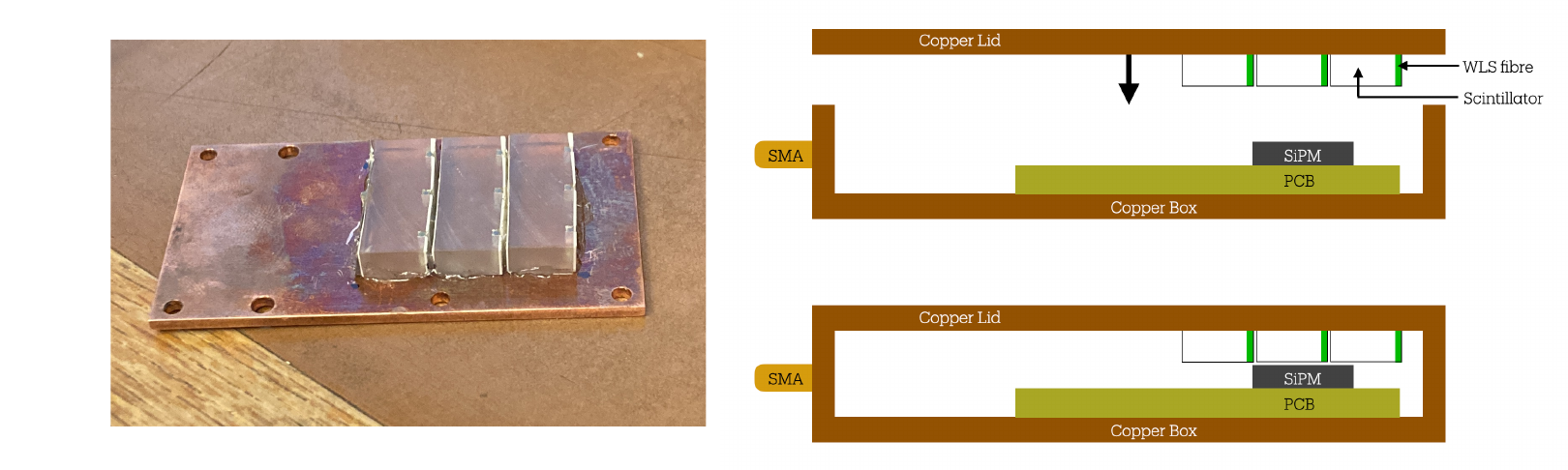}
\caption{Left: Image showing the three scintillator blocks attached to the inside of the spare copper box lid. Each block is embedded with three optical fibres, which guide photons generated in the scintillator towards the SiPM surface. Right: Cross-sectional diagram showing integration of scintillator blocks inside copper box when lid is attached and box is closed.}
\label{F:scintillator}
\qquad
\end{figure}

Three small blocks of the scintillator, each with dimensions of 1 cm (W) $\times$ 3 cm (L) $\times$ 0.5 cm (D), are attached to the inside of a spare copper box lid using epoxy adhesive. The scintillator blocks are each embedded with three wavelength shifting fibres~\cite{WLSfibres}, which shift blue light to green with absorption at 420 nm and emission peaking at $\sim$494 nm~\cite{abreu2018optimisation}, and guide the photons produced in the scintillator towards the face of the SiPM. Figure~\ref{F:scintillator} shows a picture of the scintillator blocks attached to the inside of the copper box lid. A cross-sectional diagram showing the integration of the scintillator inside the copper box when the lid is attached is shown in Figure~\ref{F:scintillator} (right). The SiPM is not physically in contact with the scintillator blocks when the box is closed; there is an extremely small gap of 1-2 mm between the scintillator blocks and the SiPM face when the box is closed. The SiPM sits in the central area of the three scintillator blocks, for maximum coverage and light collection efficiency.

\begin{figure}[htpb]
\centering
\hspace*{-1cm}  
\includegraphics[width=1.1\textwidth]{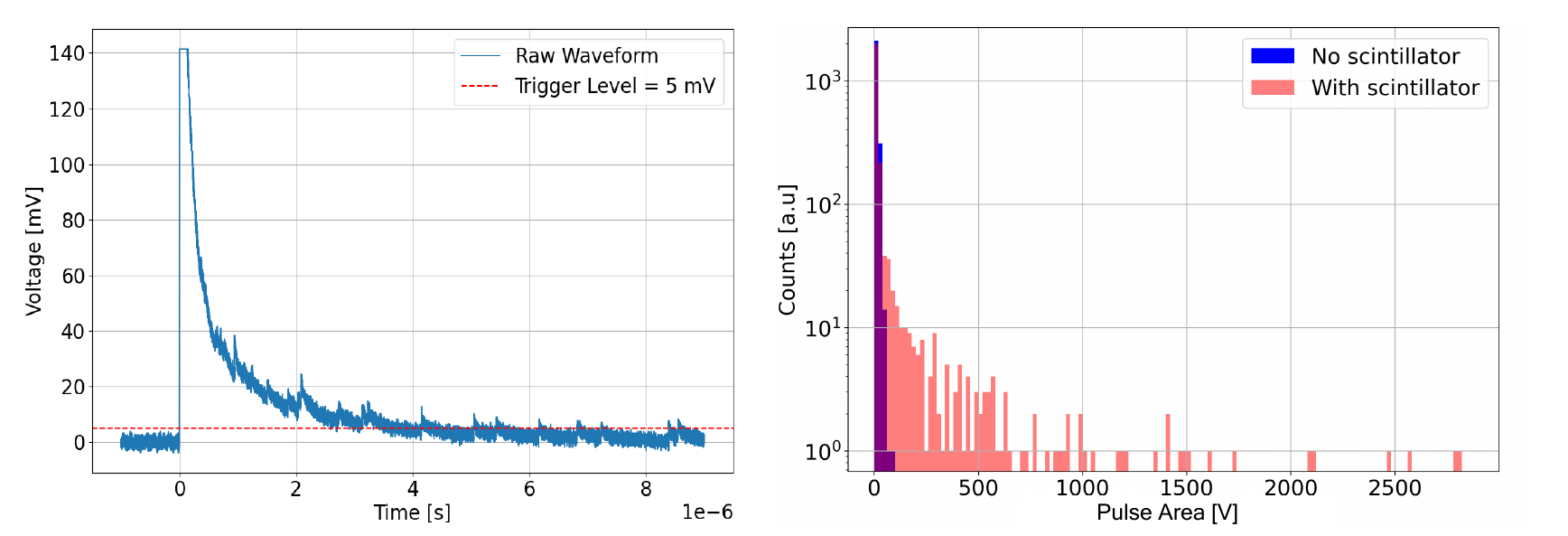}
\caption{Left: Example waveform trace taken at $V_{\mathrm{bias}}$ = 32 V ($\Delta V$ = 6.3 V), illustrating a high-light event consistent with through-going charged particles, potentially from a cosmic-ray muon. Right: Pulse area distribution (integrated over a 10 \textmu s window) about the event trigger at $V_{\mathrm{bias}}$ = 32 V ($\Delta V$ = 6.3 V) when SiPM is not coupled to scintillator (blue) versus when SiPM is coupled to scintillator (red).}
\label{F:cosmics}
\qquad
\end{figure}

Figure~\ref{F:cosmics} (left) shows an example high-light event observed in this dataset, acquired at $V_{\mathrm{bias}}$ = 32 V ($\Delta V$ = 6.3 V). In this event, the number of photons incident on the SiPM is so large that it surpasses the dynamic range of the oscilloscope, illustrated by the clipping of the signal at 140 mV. For high photon fluxes such as this, we also observe a lengthening of the signal tail, compared to the SiPM recharge time ($<$1 \textmu s). This is caused by the superposition of pulses from real photons and correlated noise, which contribute to an overall increase in the duration of the signal. 

Figure~\ref{F:cosmics} (right) compares the pulse area distribution, computed as the 10 \textmu s integral about the event trigger, for data acquired with the scintillator (red) versus data acquired without the scintillator (blue). Without scintillator, we only expect to observe signals associated with the SiPM's intrinsic noise, resulting in a narrow distribution peaked at the single photon integral. However, once scintillator is introduced, we observe a much broader distribution and a high charge tail, corresponding to energy depositions in the scintillator from particle interactions. This distribution is likely a combination of environmental gamma-rays, which dominate the lower part of the charge distribution, and high-energy cosmic-ray muons populating the high charge tail, as observed in the cryogenic cosmic-ray muon veto system developed by the NUCLEUS experiment~\cite{NUCLEUS}. However, this has not been fully characterised here. This data demonstrates that the SiPM is clearly detecting light from the scintillator and optical fibres, which is easily distinguishable above the SiPM noise sources.

As discussed in Section~\ref{subsec:APtrains}, the SiPM exhibits an increase in noise from long-lasting APs when operated at ultra-low temperatures. Whilst this additional noise source clearly limits the photon counting capability of the SiPM for a low photon flux, this is less of a concern for high photon fluxes generated from cosmic-ray muons. An increase in long-lasting APs will manifest as an increase to the integrated charge of a signal, but only in response to a primary energy deposition in the scintillator. As such, this should not limit the ability of the SiPM to tag cosmic-ray muons through their initial energy deposition. This is a first proof-of-concept of using scintillator coupled to a NUV-HD-cryo SiPM in a mK environment as a means of identifying high-light events consistent with through-going charged
particles, such as from cosmic-ray muons.

\section{Conclusions} 
\label{sec:summary}
We have shown that a FBK NUV-HD-cryo SiPM is capable of operating successfully inside a cryogen-free dilution refrigerator down to a base temperature of 9.4 $\pm$ 0.2 mK. No measurable increase in the dilution refrigerator base temperature was observed during SiPM operation, indicating that the associated power dissipation is compatible with the available thermal budget. This confirms that the device satisfies the cryogenic cosmic-ray muon veto system requirement for operation directly at the mK stage, coupled to the scintillator, without compromising the fridge performance. We measure the breakdown voltage and gain of the SiPM, and characterise several noise properties including the dark count rate, direct crosstalk probability, and correlated delayed avalanche probability, which we predominately attribute to afterpulsing. 

We also present early measurements of deploying the SiPM coupled to scintillator with optical fibre assembly inside the dilution fridge and operated at 9.4 $\pm$ 0.2 mK, as proof-of-concept for a cryogenic cosmic-ray muon veto system to be operated internal to a dilution refrigerator. In this run, we observe a significantly broader charge spectrum, consistent with high-light events originating from energy depositions from through-going charged particles, likely from environmental gamma-rays and high-energy cosmic-ray muons, which are not present in the data acquired without the scintillator. 

The SiPM used in this work originates from an engineering run used for pre-production assembly for the DarkSide-20k experiment and was developed to meet the stringent requirements of single photon detection in a near background-free environment. In contrast, the scintillator-based cryogenic cosmic-ray muon veto system we propose for the QUEST-DMC experiment will operate in a regime of much larger signals, since muons traversing the scintillator will deposit significant energy and produce many photons. The requirements on the noise levels are therefore less stringent than the expected values outlined in~\cite{NUV-HD-cryo}, provided that muon signals remain sufficiently larger than background contributions from environmental gamma interactions and intrinsic SiPM noise. The system must be able to achieve at least a 90\% muon veto efficiency, whilst minimising the experimental dead time due to accidental vetoes.

The dark count rate and direct crosstalk probability we measure in this work appear to be roughly consistent with the expected performance at 77 K for similar $\Delta V$ values~\cite{NUV-HD-cryo}, which would satisfy the noise requirements for the cosmic-ray muon veto system. These results indicate a weak temperature dependence of these properties between 77 K and 9.4 mK, however this would need to be confirmed with additional measurements at more temperature points across multiple devices. We do however observe a significant increase in the afterpulsing probability, measuring 42\% at $\Delta V$ = 6.3 V assuming a 10 \textmu s integration window, compared to an expected afterpulsing probability of 12\% at $\Delta V$ = 5 V at 77 K~\cite{NUV-HD-cryo}. In addition, we observe evidence of long-lasting afterpulse trains which can last for up to approximately 1 ms, a phenomena which has previously been observed in SiPMs at temperatures below 40 K~\cite{hanski2025performance,zhang2022scintillation}.

While this additional noise does not prevent operation of the veto system, it introduces event-by-event fluctuations in the reconstructed charge and can produce high-charge tails. This increases the probability that background events from environmental gamma interactions exceed the veto threshold, leading to a higher rate of accidental vetoes and hence increased dead time, unless a higher threshold is applied. However, raising the threshold may compromise the ability of the system to achieve the target veto efficiency of 90\%. To minimise this excess noise, it will be important to ensure that the scintillation signal is sufficiently prompt, such that the primary light is collected within a short time window. This consideration will inform the next-stage of R\&D of the cosmic-ray muon veto system, in particular in the choice of scintillator material, and in optimisation of the geometry and light collection (including the number of required SiPMs and the scintillator-SiPM coupling), to achieve a fast and well-defined muon signal. Following this, we will characterise the cryogenic cosmic-ray muon veto system as a function of temperature, from room temperature down to mK temperatures, in order to determine the minimum veto threshold required to achieve the target performance for the QUEST-DMC experiment.

\acknowledgments
This work was funded by UKRI EPSRC and STFC (Grants ST/T006773/1, ST/Y004434/1, EP/P024203/1, EP/W015730/1 and EP/W028417/1), as well as the European Union’s Horizon 2020 Research and Innovation Programme under Grant Agreement no 824109 (European Microkelvin Platform). S.A. acknowledges financial support from the Jenny and Antti Wihuri Foundation. M.D.T acknowledges financial support from the Royal Academy of Engineering (RF/201819/18/2). J.Sm. acknowledges support from the UK Research and Innovation Future Leader Fellowship MR/Y018656/1. A.K. acknowledges support from the UK Research and Innovation Future Leader Fellowship MR/Y019032/1.

\bibliographystyle{JHEP}
\bibliography{biblio.bib}

\end{document}